\documentclass[fleqn,usenatbib]{mnras}
\usepackage{newtxtext,newtxmath}
\usepackage[T1]{fontenc}
\DeclareRobustCommand{\VAN}[3]{#2}
\let\VANthebibliography\thebibliography
\def\thebibliography{\DeclareRobustCommand{\VAN}[3]{##3}\VANthebibliography}
\usepackage{graphicx}
\usepackage{amsmath}
\usepackage{xcolor}
\usepackage{orcidlink}

\newcommand{\msolar}{M$_{\odot}$}

\title[JWST/MIRI detection of SN 1980K]{Serendipitous detection of the dusty Type IIL SN 1980K with JWST/MIRI}

\author[Sz. Zsíros et al.]{Szanna Zsíros\orcidlink{0000-0001-7473-4208},$^{1}$\thanks{E-mail: szannazsiros@titan.physx.u-szeged.hu}
Tam\'as Szalai\orcidlink{0000-0003-4610-1117},$^{1,2}$
Ilse De Looze,$^{3}$
Arkaprabha Sarangi\orcidlink{0000-0002-9820-679X},$^{4}$
Melissa Shahbandeh\orcidlink{0000-0002-9301-5302},$^{5,6}$
\newauthor Ori D. Fox\orcidlink{0000-0003-2238-1572},$^{6}$
Tea Temim\orcidlink{0000-0001-7380-3144},$^{7}$
Dan Milisavljevic\orcidlink{0000-0002-0763-3885},$^{8,9}$
Schuyler D.~Van Dyk\orcidlink{0000-0001-9038-9950},$^{10}$
Nathan Smith\orcidlink{0000-0001-5510-2424},$^{11}$ 
\newauthor Alexei V. Filippenko\orcidlink{0000-0003-3460-0103},$^{12}$
Thomas G. Brink\orcidlink{0000-0001-5955-2502},$^{12}$
WeiKang Zheng,$^{12}$
Luc Dessart,$^{13}$
Jacob Jencson\orcidlink{0000-0001-5754-4007},$^{5}$
\newauthor Joel Johansson\orcidlink{0000-0001-5975-290X},$^{14}$ 
Justin Pierel,$^{6}$ 
Armin Rest\orcidlink{0000-0002-4410-5387},$^{5,6}$
Samaporn Tinyanont\orcidlink{0000-0002-1481-4676},$^{15,16}$
\newauthor Maria Niculescu-Duvaz,$^{17}$
M.~J. Barlow\orcidlink{0000-0002-3875-1171},$^{17}$
Roger Wesson,$^{18}$
Jennifer Andrews,$^{19}$ 
Geoff Clayton,$^{20}$ 
\newauthor Kishalay De,$^{21}$ 
Eli Dwek,$^{22}$ 
Michael Engesser,$^{6}$
Ryan J. Foley,$^{15}$
Suvi Gezari,$^{6}$ 
Sebastian Gomez\orcidlink{0000-0001-6395-6702},$^{6}$ 
\newauthor Shireen Gonzaga,$^{6}$ 
Mansi Kasliwal,$^{23}$
Ryan Lau,$^{24}$ 
Anthony Marston,$^{25}$ 
Richard O'Steen,$^{6}$
\newauthor Matthew Siebert,$^{6}$ 
Michael Skrutskie,$^{26}$ 
Lou Strolger,$^{6}$ 
Qinan Wang,$^{5}$
Brian Williams,$^{22}$ 
\newauthor Robert Williams,$^{6}$ 
Lin Xiao$^{27,28}$
\\
\\
$^{1}$Department of Experimental Physics, Institute of Physics, University of Szeged, H-6720 Szeged, Dóm tér 9, Hungary\\
$^{2}$HUN-REN--SZTE Stellar Astrophysics Research Group, H-6500 Baja, Szegedi {\'u}t, Kt. 766, Hungary \\
$^{3}$Sterrenkundig Observatorium, Ghent University, Krijgslaan 281 -- S9, 9000 Gent, Belgium\\
$^{4}$DARK, Niels Bohr Institute, University of Copenhagen, Jagtvej 128, 2200 Copenhagen, Denmark\\
$^{5}$Department of Physics and Astronomy, Johns Hopkins University, Baltimore, MD 21218, USA\\
$^{6}$Space Telescope Science Institute, 3700 San Martin Drive, Baltimore, MD 21218, USA\\
$^{7}$Department of Astrophysical Sciences, Princeton University, Princeton, NJ 08544, USA\\
$^{8}$Purdue University, Department of Physics and Astronomy, 525 Northwestern Ave, West Lafayette, IN 47907, USA\\
$^{9}$Integrative Data Science Initiative, Purdue University, West Lafayette, IN 47907, USA\\
$^{10}$Caltech/IPAC, Mailcode 100-22, Pasadena, CA 91125, USA\\
$^{11}$Steward Observatory, University of Arizona, 933 N. Cherry St, Tucson, AZ 85721, USA\\
$^{12}$Department of Astronomy, University of California, Berkeley, CA 94720-3411, USA\\
$^{13}$Institut d’Astrophysique de Paris, CNRS–Sorbonne Université, 98 bis boulevard Arago, F-75014 Paris, France\\
$^{14}$Oskar Klein Centre, Department of Physics, Stockholm University, AlbaNova, SE-10691 Stockholm, Sweden\\
$^{15}$Department of Astronomy and Astrophysics, University of California, Santa Cruz, CA 95064, USA\\
$^{16}$National Astronomical Research Institute of Thailand, 260 Moo 4, Donkaew, Maerim, Chiang Mai, 50180, Thailand\\
$^{17}$Department of Physics \& Astronomy, University College London, Gower St, London WC1E 6BT, UK\\
$^{18}$School of Physics and Astronomy, Cardiff University, Queen’s Buildings, The Parade, Cardiff, CF24 3AA, UK\\
$^{19}$Gemini Observatory, 670 N. Aohoku Place, Hilo, Hawaii, 96720, USA\\
$^{20}$Physics \& Astronomy, Louisiana State University, Baton Rouge, LA, USA\\
$^{21}$MIT-Kavli Institute for Astrophysics and Space Research, 77 Massachusetts Ave., Cambridge, MA 02139, USA\\
$^{22}$Observational Cosmology Lab, NASA Goddard Space Flight Center, Code 665, Greenbelt, MD 20771, USA\\
$^{23}$Cahill Center for Astrophysics, California Institute of Technology, 1200 E. California Blvd. Pasadena, CA 91125, USA\\
$^{24}$NSF’s NOIRLab, 950 N. Cherry Avenue, Tucson, 85719, AZ, USA\\
$^{25}$European Space Agency (ESA), ESAC, 28692 Villanueva de la Canada, Madrid, Spain\\
$^{26}$Department of Astronomy, University of Virginia, Charlottesville, VA 22904-4325, USA\\
$^{27}$Department of Physics, College of Physical Sciences and Technology, Hebei University, Wusidong Road 180, Baoding 071002, China\\
$^{28}$Key Laboratory of High-precision Computation and Application of Quantum Field Theory of Hebei Province, Hebei University, Baoding 071002, China
} 

\date{Accepted XXX. Received YYY; in original form ZZZ}

\pubyear{2023}

\begin{document}
\label{firstpage}
\pagerange{\pageref{firstpage}--\pageref{lastpage}}
\maketitle

\clearpage
\begin{abstract}

We present mid-infrared (mid-IR) imaging of the Type IIL supernova (SN) 1980K with the {\it James Webb Space Telescope} ({\it JWST}) more than 40 yr post-explosion. SN~1980K, located in the nearby ($D\approx7$~Mpc) ``SN factory'' galaxy NGC 6946, was serendipitously captured in {\it JWST}/MIRI images taken of the field of SN~2004et in the same galaxy. SN 1980K serves as a promising candidate for studying the transitional phase between young SNe and older SN remnants and also provides a great opportunity to investigate its the close environment.
SN~1980K can be identified as a clear and bright point source in all eight MIRI filters from F560W up to F2550W.
We fit analytical dust models to the mid-IR spectral energy distribution that reveal a large amount ($M_d \approx 0.002~{\rm M}_{\odot}$) of Si-dominated dust at $T_\rmn{dust}\approx 150$~K (accompanied by a hotter dust/gas component), and also computed numerical SED dust models. 
Radiative transfer modeling of a late-time optical spectrum obtained recently with Keck discloses that an even larger ( $\sim 0.24-0.58$ \msolar) amount of dust is needed in order for selective extinction to explain the asymmetric line profile shapes observed in SN~1980K. 
As a conclusion, with {\it JWST}, we may see i) pre-existing circumstellar dust heated collisionally (or, partly radiatively), analogous to the equatorial ring of SN~1987A, or ii) the mid-IR component of the presumed newly-formed dust, accompanied by much more colder dust present in the ejecta (as suggested by the late-time the optical spectra).

\end{abstract}

\begin{keywords}
infrared: stars -- supernovae: general -- supernovae: individual: SN~1980K -- dust, extinction
\end{keywords}



\section{Introduction}


Core-collapse supernovae (CCSNe), the cataclysmic endings of evolved massive stars, are unique astrophysical laboratories. 
Due to their particularly energetic final explosions, they affect their closer and broader environments and enable us to uncover details about the pre-explosion stellar evolution processes.

CCSNe have long been considered as crucial sources of dust in the Universe, potentially accounting for the origin of dust at high redshifts during the epoch of reionisation \citep{maiolino04,dwek07,gall11}. 
Recent far-infrared (far-IR) and sub-mm observations of Galactic SN remnants -- e.g., Cas~A \citep{barlow10,sibthorpe10,arendt14} and Crab \citep{gomez12,temim13,delooze19} -- and the very nearby ($\sim 50$~kpc) SN~1987A \citep{matsuura11,matsuura19,indebetouw14} seem to confirm that massive ($\sim0.1$--1~\msolar) cold reservoirs may be hiding significant quantities of dust. 
Such long-wavelength studies, however, cannot be carried out for other extragalactic CCSNe owing to the limited sensitivity of previous far-IR instruments. 
Dust masses estimated by follow-up {\it Spitzer Space Telescope} (hereafter {\it Spitzer}) observations resulted in 2--3 orders of magnitude smaller dust masses; nevertheless, these data tended to probe only the warmer dust components ($>500$~K) and/or earlier epochs ($<5$~yr). 
In addition to their thermal radiation, dust grains in the SN ejecta preferentially absorb light in the receding part of the ejecta, attenuating the red wing of emission lines at optical wavelengths. 
Applying this method, several authors \citep{bevan16,bevan17,niculescu-duvaz22} also found large  ($>10^{-2}$~\msolar) dust masses in years- or decades-old CCSNe; however, this kind of analysis gives information only on the mass of dust, but not on its temperature.

Observed dust in CCSNe may form either in the (unshocked) ejecta or in a cold dense shell (CDS) across the contact discontinuity between the shocked circumstellar matter (CSM) and shocked ejecta \citep[see, e.g.,][]{chugai04,pozzo04,smith08,smith09,mattila08}. 
A late-time mid-IR excess may also emerge from heated pre-existing dust grains. 
While in the shocked CSM, heating can be collisional, and grains in the more distant, unshocked CSM are assumed to be radiatively heated by the peak SN luminosity or by energetic photons generated during CSM interaction, thereby forming an IR echo \citep[see, e.g.,][]{bode80,dwek83,graham86,sugerman03,smith08b,kotak09,fox10,andrews11}. 
In these cases, dust can be a helpful probe of the CSM characteristics and the pre-SN mass loss from either the progenitor or companion star \citep[see, e.g.,][for a review]{gall11}.

Multiwavelength follow-up observations of the evolution of the expanding SN ejecta and their interaction with the ambient medium plays an essential role in this research.
Beyond gathering optical and near-IR data using ground-based infrastructure, mid-IR observations offer numerous advantages for following the late-time evolution of SNe owing to the increased sensitivity to the expanding and cooling ejecta and the lower impact of interstellar extinction. 
This wavelength region also covers atomic and molecular emission lines generated by shocked, cooling gas \citep{reach06}. 
Moreover, mid-IR data are sensitive to warm dust either in the SN ejecta or in the pre-existing CSM.

In the last two decades, the prime source of mid-IR SN data was NASA's now-decommissioned {\it Spitzer}, which provided valuable data during both its cryogenic (2003--2009) and post-cryogenic (2009-2020) missions. 
Beyond a few large-scale SN surveys --- e.g., the SPIRITS project \citep[SPitzer InfraRed Intensive Transients Survey, a systematic study of transients in nearby galaxies; see][]{tinyanont16,kasliwal17,jencson19}, or other studies focused on CSM-interacting (Type IIn) SNe \citep{fox11,fox13} --- and several single-object studies, many other objects appeared in nontargeted archival {\it Spitzer} images. \citet{szalai19,szalai21} presented the most extensive analyses of mid-IR SN data, including $\sim 120$ positively detected objects from $\sim 1100$ SN sites imaged by {\it Spitzer}.

These latter studies focus primarily on the comprehensive examination of 3.6~$\mu$m and 4.5~$\mu$m photometric datasets of SNe obtained with the InfraRed Array Camera (IRAC) detector of {\it Spitzer}. A limited number of longer-wavelength measurements were obtained during the cryogenic mission with IRAC (5.8~$\mu$m and 8.0~$\mu$m), the Multiband Imaging Photometer (MIPS; 24~$\mu$m), and the InfraRed Spectrograph for Spitzer (IRS; $\sim 5$--16~$\mu$m). 
Detailed works based on extended mid-IR datasets have been published for several Type IIP SNe \citep[e.g.,][]{kotak09,fabbri11,meikle11,szalai11,szalai13}, but also Type IIn SN~1978K \citep{tanaka12}, Type IIb SN~1993J \citep{zsiros22}, and the famous, peculiar Type II SN~1987A \citep{bouchet06,dwek10,arendt16,arendt20}.

The {\it James Webb Space Telescope (JWST)} offers a new opportunity to detect the late phases of cool ($\sim100$--200~K) dust in extragalactic SNe beyond SN 1987A. 
{\it JWST} offers unique sensitivity to (i) cooler dust grains at wavelengths $>4.5$~$\mu$m, (ii) the 10~$\mu$m silicate feature that can distinguish grain compositions, and (iii) faint emission from the SN at very late ($>2000$~days) epochs that would have gone undetected by {\it Spitzer} and any other mid-IR spacecraft.
In its first half year, the new space telescope has begun to revolutionise this field. 
During Cycle 1 General Observers (GO) program 2666 (PI O. Fox; DOI: \doi[10.17909/8kkm-fr55]{http://dx.doi.org/10.17909/8kkm-fr55}), {\it JWST} has already detected dust in SNe~IIP 2004et and 2017eaw. 
In SN~2004et, the observations have uncovered the largest newly-formed ejecta dust masses in an extragalactic SN other than SN~1987A, with $\gtrsim10^{-2}$~\msolar\ dust residing at $\sim 140$~K \citep{Shahbandeh2023}. 

Here we present the serendipitous detection of another CCSN, SN~1980K, with {\it JWST} during the analysis of images obtained originally for SN~2004et. 
This paper is organised as follows. 
In Section~\ref{sec:obs}, we present our observations and data reduction. 
Section~\ref{sec:anal} describes the steps of measuring dust masses via modeling the mid-IR spectral energy distribution (SED) and the red-blue line-profile asymmetries in a newly-obtained late-time optical spectrum. 
We interpret the results in Section~\ref{sec:disc}, discuss the possible origin and heating mechanisms of dust, and provide our concluding remarks.

\section{Observations and data reduction}\label{sec:obs}

\subsection{SN~1980K}\label{sec:obs_80K}

SN~1980K, discovered on 1980 Oct. 28 (UTC dates are used throughout this paper) in the nearby ($D \approx 7.12$~Mpc) ``SN factory'' galaxy NGC~6946 \citep{wild80}, has long been a target of multiwavelength studies. 
Based on the early-time photometric and spectroscopic observations showing the presence of medium-strong H lines and a linearly (in magnitudes) declining light curve, it was classified as a prototype Type II-linear (IIL) SN \citep{buta82,barbon82}. 
However, note that despite some obvious photometric and spectral differences, there are doubts on whether Type IIL and more slowly-evolved, H-rich Type II-plateau (IIP) SNe truly originate from different types of progenitors \citep[see, e.g.,][]{Anderson_2014,Valenti_2016}.
Regarding SN~1980K, there is no direct information on the progenitor star (or stellar system). 
Based on the nondetection of the progenitor on a pre-explosion image, \cite{Thompson_1982} determined an upper mass limit of 18~\msolar, while \cite{Williams_2018} and \cite{Koplitz_2021} found a progenitor mass range of 7--15~\msolar\ (giving 7.5--9~\msolar\ as the most probable value) from an analysis of the local stellar population.

About 2 months after the discovery, a narrow H$\alpha$ emission line with a weak blueshifted absorption component appeared in the spectrum of SN~1980K \citep{barbieri82}, indicating an ongoing SN shock-CSM interaction (just as revealed in the case of SN~1979C, another SN~IIL discovered a year before SN~1980K). 
Long-term follow-up spectroscopy of SN~1980K showed only a small ($\sim 25$\%) decline of the H$\alpha$ flux over a decade \citep{uomoto86,leibundgut91,fesen95,fesen99}. 
Spectra obtained $\sim 15$~yr and 30~yr after the explosion also revealed the presence of broad ($\sim 5500$~km~s$^{-1}$) emission lines of [O~{\sc i}] and [O~{\sc ii}] \citep{milisavljevic12}. 
Similar findings were presented by \cite{Long19} in the case of a Gemini Multi-Object Spectrograph (GMOS) spectrum taken $\sim 34$~yr post-explosion, confirming the slow decline of the H$\alpha$ flux at later times.
The slow rate of change of spectral features --- as well as the presence of late-time IR emission --- can be also explained by scattered and thermal light echoes from extended circumstellar or interstellar matter \citep[][see details later]{dwek83,sugerman12}.

As further evidence of CSM interaction, radio and X-ray emission were detected from SN~1980K about a month after maximum optical brightness \citep{canizares82,weiler86}. 
Radio observations of SN~1980K were continued extensively in the following years \citep{weiler92,montes98,eck02}, resulting in positive detections but showing declining fluxes. 
However, as discussed by \cite{montes98}, the sharp drop in radio emission seemed to indicate a notable change in the CSM density, with the shock presumably entering a different CSM regime.
In the X-ray, a more limited dataset exists; however, it is worth highlighting that the SN was detected by {\it ROSAT} \citep{schlegel94} and {\it Chandra} \citep{soria08,fridriksson08} more than 10 and 20~yr after the explosion, respectively.

\subsection{Mid-IR photometry on {\it JWST/MIRI} images}

As part of GO-2666,we obtained images of NGC~6946 with the {\it JWST} Mid-Infrared Instrument (MIRI; \citealt{Bouchet_2015, Ressler_2015, Rieke_2015, Rieke_2022}) on 2022~Sep.~20.9, originally targeting Type IIP SN~2004et.
The observations were acquired in the F560W, F1000W, F1130W, F1280W, F1500W, F1800W, F2100W, and F2550W filter bands, using the FASTR1 readout pattern in the FULL array mode and a 4-point extended source dither pattern.
The detailed calibration process of the {\it JWST}/MIRI images, together with the analysis of SN~2004et, has been recently published by \citet{Shahbandeh2023}.

\begin{figure*}
        \includegraphics[width=\textwidth]{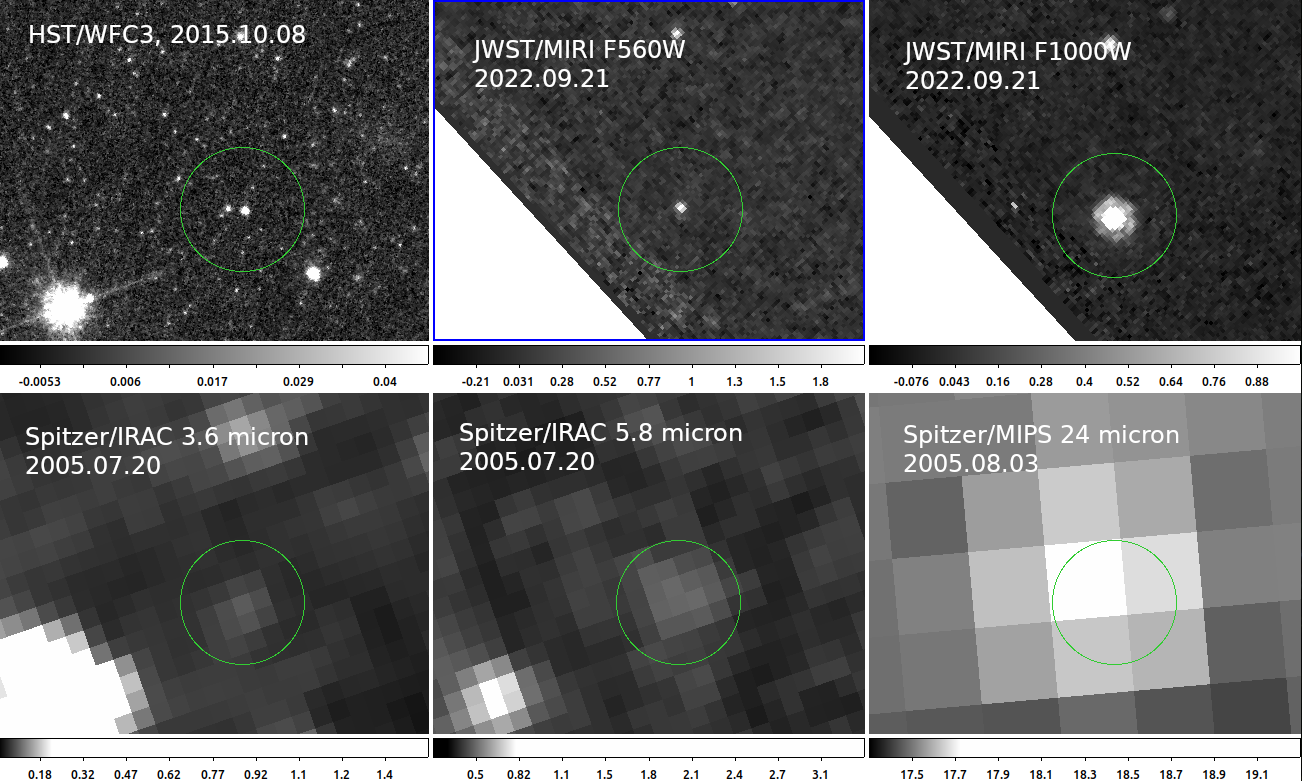}
    \caption{SN~1980K in {\it JWST}/MIRI F560W and F1000W images compared with archival {\it HST}/WFC3 and {\it Spitzer}/IRAC and MIPS images of the field (north is up and east is to the left). Green circles with a radius of 2\arcsec\ are centred on the position of the SN in {\it HST}/WFC3 images.} 
    \label{fig:80K_img1}
\end{figure*}

\begin{figure*}
        \includegraphics[width=\textwidth]{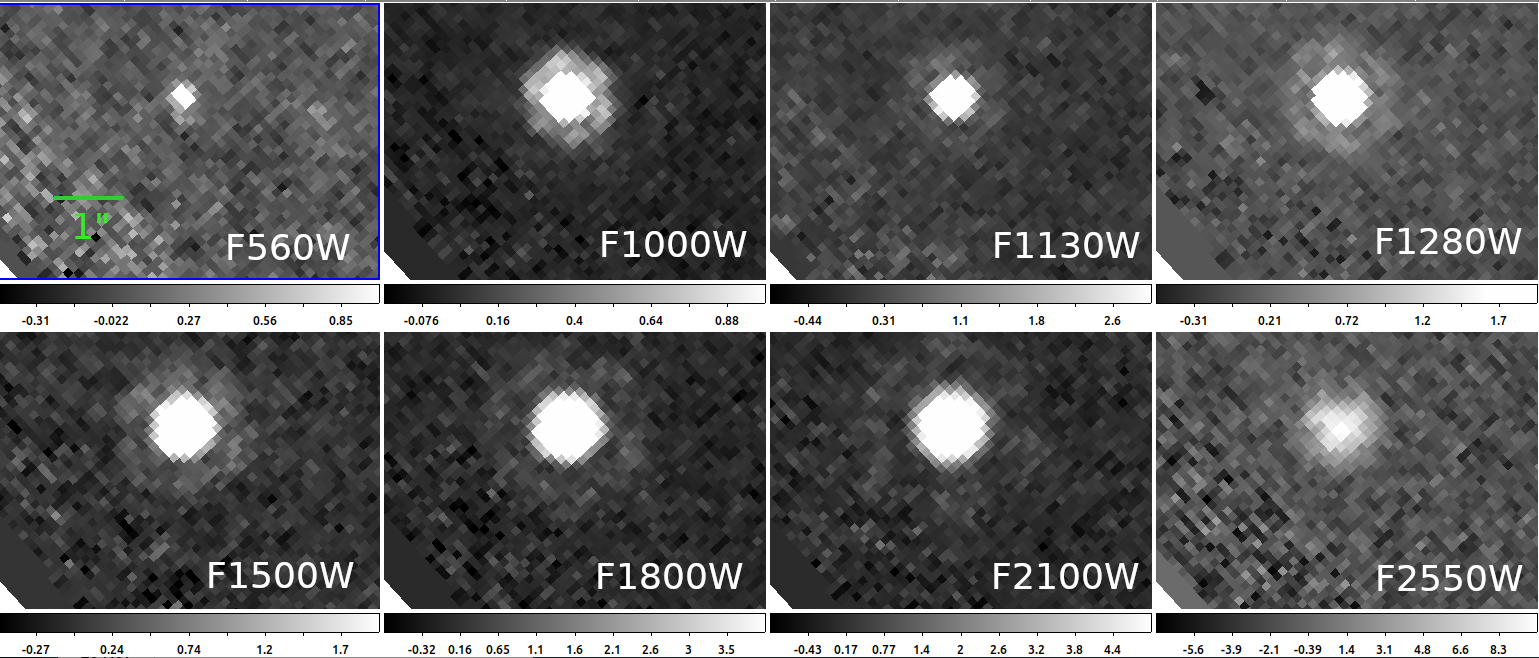}
    \caption{SN~1980K on all background-subtracted {\it JWST}/MIRI images taken on 2022 Sep. 20.9 of the field of SN~2004et in NGC~6946.}
    \label{fig:80K_img2}
\end{figure*}

Serendipitously, the MIRI field of view also contains the site of SN~1980K \citep[$\alpha(2000) = 20^{\rm h}35^{\rm m}30.07^{\rm s}$, $\delta(2000) = +60^{\circ} 06\arcmin 23\farcs 8$;][]{vandyk96}. 
The object can be identified as a clear and bright point source at all wavelengths from 5.6 to 25.5~$\mu$m (see in Figs. \ref{fig:80K_img1} and \ref{fig:80K_img2}), for which high-resolution {\it Hubble Space Telescope (HST)} images --- obtained in 2008 January showing the environment of SN 1980K \citep{milisavljevic12} --- proved to be highly useful.
We used the JWST HST Alignment Tool \citep[JHAT,][]{Rest_2023} to align the {\it JWST} and {\it HST} images to each other.

\begin{figure}
        \includegraphics[width=0.45\textwidth]{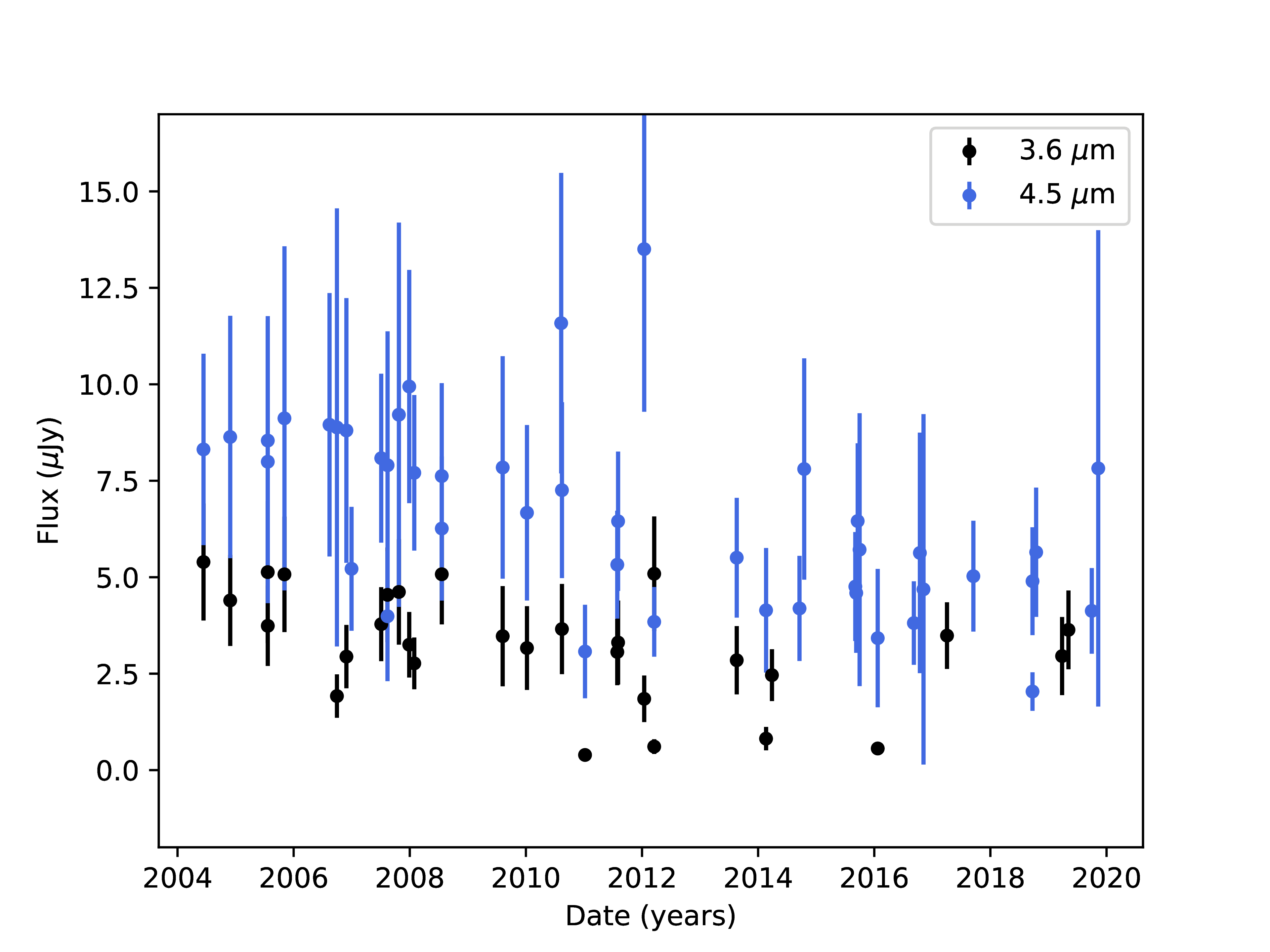}
    \caption{{\it Spitzer}/IRAC fluxes of SN~1980K in the 3.6~$\mu$m and 4.5~$\mu$m channels. The SN does not show significant evolution over the course of 15~yr of {\it Spitzer} images.}
    \label{fig:80K_spitzer}
\end{figure}

To measure the fluxes of SN~1980K on {\it JWST}/MIRI images, we followed the method described in detail by \citet{Shahbandeh2023}. 
We performed point-spread-function (PSF) photometry on background-subtracted level-two data products using \texttt{WebbPSF} \citep{Perrin_2014}.
In order to calibrate the flux, we applied flux offsets by measuring the PSF of all the stars in the field and comparing them to the corresponding catalogues created by the pipeline. 
The fluxes of all four dithers of each filter were then averaged. 
The final results of {\it JWST}/MIRI photometry of SN~1980K are presented in Table~\ref{tab:JWST_phot}.

\begin{table}
	\centering
	\caption{Mid-IR fluxes of SN~1980K.$^a$}
	\label{tab:JWST_phot}
    \renewcommand{\arraystretch}{1.5}
	\begin{tabular}{lc} 
		\hline
		$\lambda$ & $F_{\nu}$\\
       ($\mu$m) &($\mu$Jy)\\
        \hline
		5.6 & 99.800 $\pm$ 0.565\\
        10.0 & 74.700 $\pm$ 0.169\\
        11.3 & 73.400 $\pm$ 0.572\\
        12.8 & 100.198 $\pm$ 0.570\\
        15.0 & 144.222 $\pm$ 0.761\\
        18.0 & 269.199 $\pm$ 1.570\\
        21.0 & 291.572 $\pm$ 2.080\\
        25.5 & 312.250 $\pm$ 6.460\\
		\hline
	\end{tabular}
 
 $^a$Obtained from {\it JWST}/MIRI images on 2022~Sep.~20.9 ($\sim 42$~yr post-explosion).
\end{table}

\subsection{Mid-IR photometry on archival {\it Spitzer} images}\label{sec:obs_spitzer}

NGC~6946, the nearby host galaxy of SN~1980K, was imaged several times with {\it Spitzer}, especially during its Warm Mission Phase \citep[as part of SPIRITS program;][]{tinyanont16}. 
Based on IRAC and MIPS data obtained between 2005 and 2008, \cite{sugerman12} carried out a detailed analysis of SN~1980K; however, the complete {\it Spitzer} dataset of the SN had not been published.
Thus, we reanalysed all of the {\it Spitzer} IRAC and MIPS data obtained between 2004 and 2019. 
We downloaded the post-basic calibrated (PBCD) images of SN~1980K from the {\it{Spitzer Heritage Archive (SHA)}}\footnote{\href{https://sha.ipac.caltech.edu/applications/Spitzer/SHA/}{https://sha.ipac.caltech.edu/applications/Spitzer/SHA/}}.  

Despite the limitations of the spatial and spectral resolution, on most of the images a faint source can be detected at the position of the SN. 
At the spatial resolution of {\it Spitzer}, the region of the SN seems to be complex and a nearby bright source (see in Fig.\ref{fig:80K_img1}) also makes single-aperture photometry difficult.
Therefore, we applied a photometric method described by \citet{fox11} that uses a set of single apertures to measure the background and source fluxes individually. 
It allows us to manually eliminate the effect of nearby bright sources and sample only the local background of the SN. 

We carried out the photometry using the {\it{phot}} task of IRAF {\it{Image Reduction and Analysis Facility)}} software package\footnote{\href{https://iraf.noao.edu}{https://iraf.noao.edu}}. 
For the IRAC and MIPS images, we used an aperture radius of 2.4\arcsec\ and 5\arcsec, respectively.
To obtain monochromatic fluxes, we applied aperture corrections: according to the IRAC\footnote{\href{https://irsa.ipac.caltech.edu/data/SPITZER/docs/irac/iracinstrumenthandbook/}{https://irsa.ipac.caltech.edu/data/SPITZER/docs/irac/iracinstrumenthandbook/}} and MIPS\footnote{\href{https://irsa.ipac.caltech.edu/data/SPITZER/docs/mips/mipsinstrumenthandbook/}{https://irsa.ipac.caltech.edu/data/SPITZER/docs/mips/mipsinstrumenthandbook/}} Instrument Handbooks, we used values of 1.213, 1.234, 1.379, and 1.584 for the IRAC 3.6, 4.5, 5.8, and 8.0~$\mu$m channels, respectively, and 2.12 for the MIPS 24.0~$\mu$m channel.

Based on our analysis, the SN does not show significant evolution over the course of 15~yr of {\it {Spitzer}} observations (see in Fig. \ref{fig:80K_spitzer}). Hence, we compared the average IRAC and MIPS fluxes with the MIRI data and found them to be similar. However, since {\it{Spitzer}} data have large uncertainties owing to their low spatial resolution compared to that of {\it JWST}/MIRI, we do not include them later in our analysis. 
Note that both the IRAC and MIPS photometry have high uncertainties (Fig.~\ref{fig:80K_spitzer}), and the obtained fluxes also depend on the applied aperture configuration. 
Thus, we only used them to determine whether the data shows a trend in time and also to estimate the late-time fluxes at shorter wavelengths (3.6$\mu$m and 4.5$\mu$m).

\subsection{Late-time optical spectrum}\label{sec:obs_optspec}

\begin{figure*}
        \includegraphics[width=\textwidth]{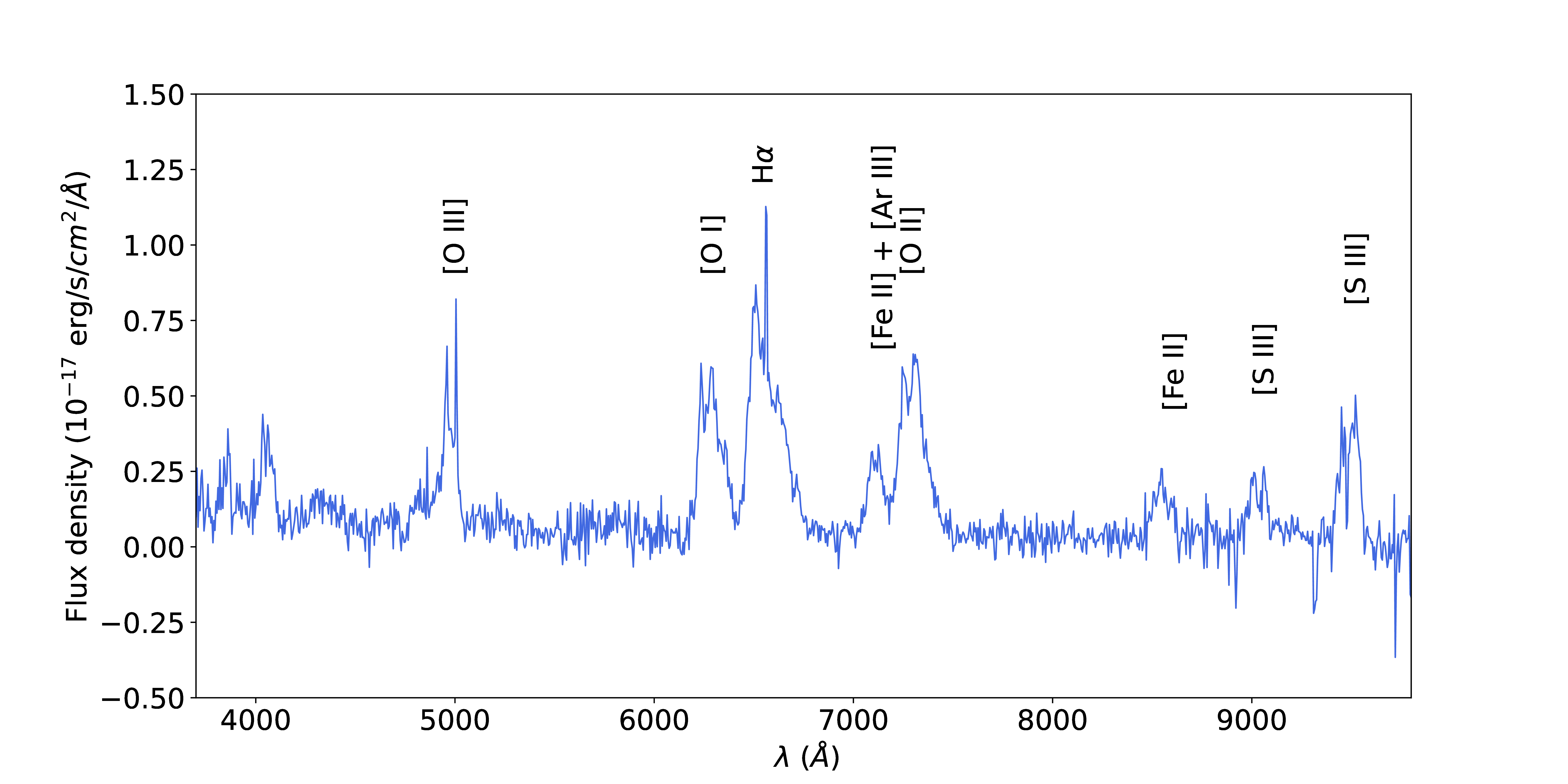}
    \caption{Late-time optical spectrum of SN~1980K obtained with Keck/LRIS on 2022~November~19, at an epoch of 15\,362~days ($\sim 42$~yr) post-explosion. The spectrum has been binned to 5 \AA\ per pixel.}
    \label{fig:80K_sp1}
\end{figure*}

A late-time optical spectrum of SN~1980K was obtained with the Keck Low Resolution Imaging Spectrometer (LRIS; \citealp{Oke_1995}) on 2022~Nov.~19 (at an epoch of $\sim 42$~yr); it is shown in Figure \ref{fig:80K_sp1}. 
The spectrum was acquired with the slit oriented at or near the parallactic angle to minimise slit losses caused by atmospheric dispersion \citep{Filippenko_1982}. 
The LRIS observations utilised the $1\arcsec$-wide slit, 600/4000 grism, and 400/8500 grating to produce a similar spectral resolving power ($R \approx 700$--1200) in the red and blue channels. 

Data reduction followed standard techniques for CCD processing and spectrum extraction using the LPipe data-reduction pipeline \citep{Perley_2019}. 
Low-order polynomial fits to comparison-lamp spectra were used to calibrate the wavelength scale, and small adjustments derived from night-sky lines in the target frames were applied. 
The spectrum was flux calibrated using observations of appropriate spectrophotometric standard stars observed on the same night, at similar airmasses, and with an identical instrument configuration.

Beyond the H$\alpha$ and [O~{\sc i}] $\lambda\lambda$6300, 6363 lines, of which we carried out a detailed profile analysis (Sect. \ref{sec:anal_optspec}), we also identified, following \cite{milisavljevic12}, [O~{\sc iii}] $\lambda$5007, [Fe~{\sc ii}] $\lambda$7155 (associated probably with [Ar~{\sc iii}]), and [O~{\sc ii}] $\lambda$7300. Moreover, owing to also covering the longer-wavelength part of the spectrum with a good signal-to-noise ratio, we also see [Fe~{\sc ii}] $\lambda$8617 and [S~{\sc iii}] $\lambda\lambda$9069, 9531 lines (Fig. \ref{fig:80K_sp1}).

\section{Analysis}\label{sec:anal}

\subsection{Analytical models of the mid-IR SED}\label{sec:anal_analysis}

\begin{figure*}
    \includegraphics[width=0.45\textwidth]{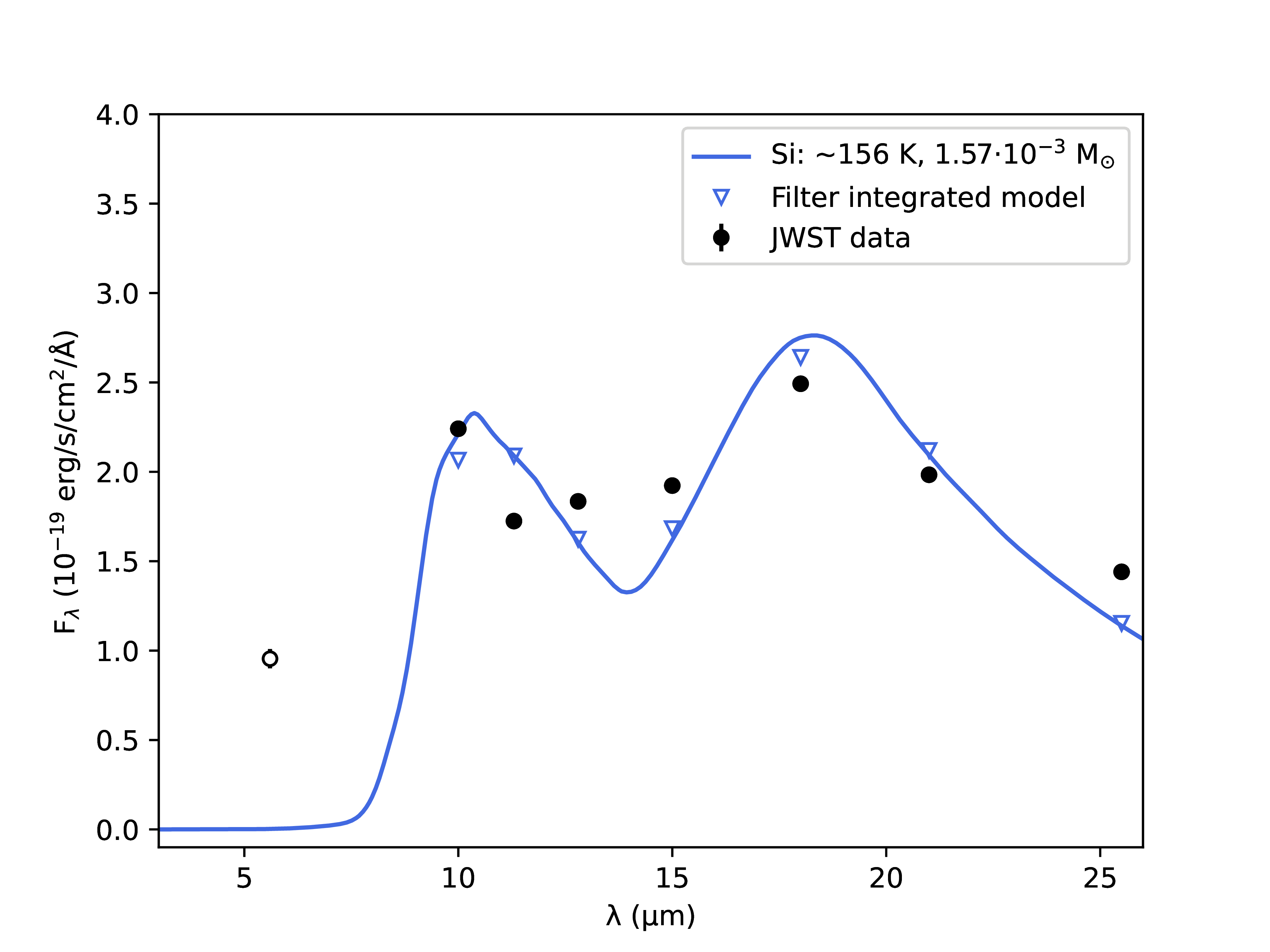}
    \includegraphics[width=0.45\textwidth]{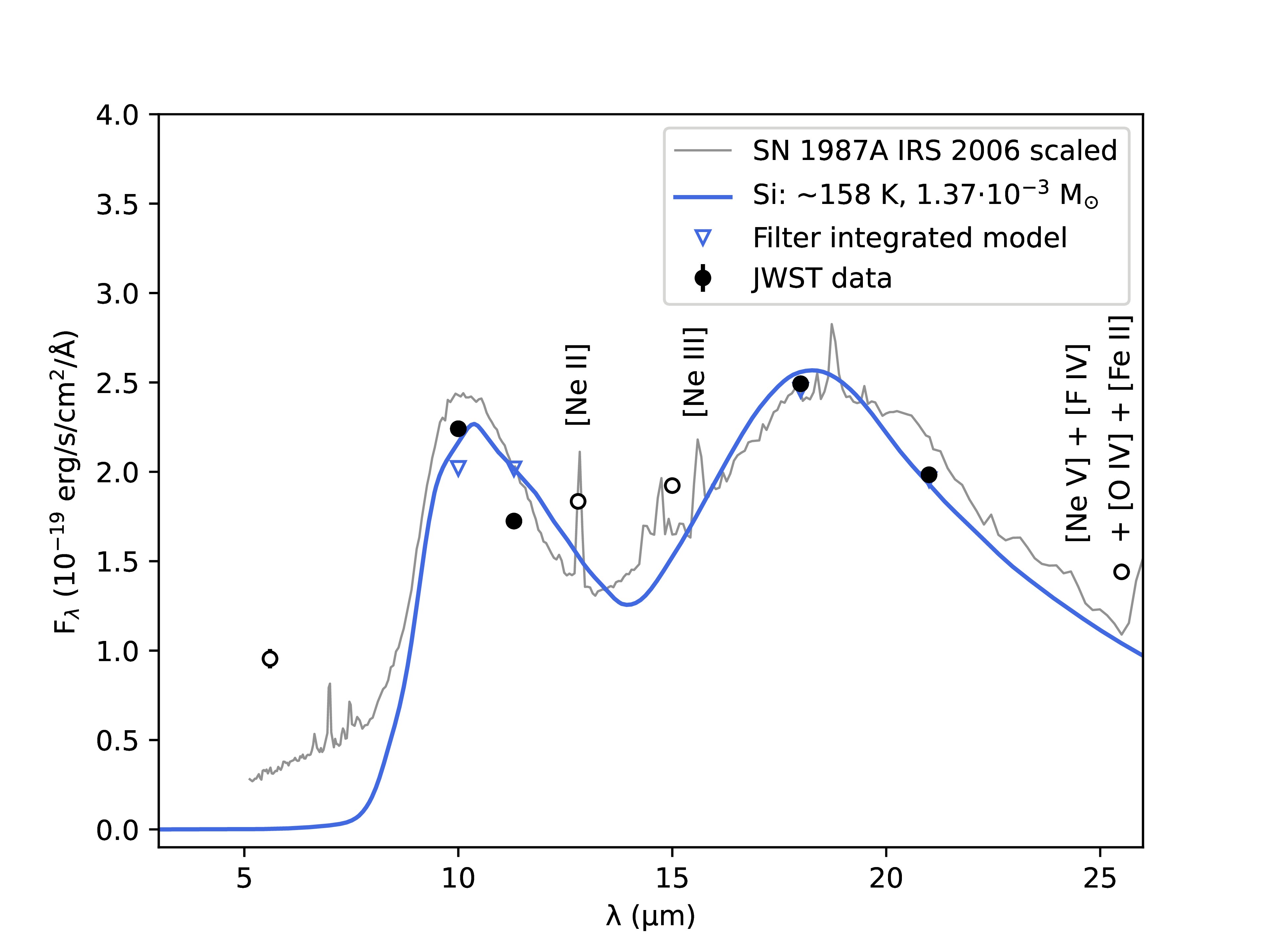}
    \caption{Best-fit one-component silicate dust models fitted to the mid-IR SED of SN~1980K excluding the 5.6~$\mu$m data point (on the left) and also the 12.8, 15.0, and 25.0~$\mu$m data points (on the right). Black dots represent the {\it JWST} data (empty circles represent the excluded data points), while blue lines represent the filter-integrated model fluxes (see details of the model-fitting process in the text). 
    On the right panel, an upscaled {\it Spitzer}/IRS spectrum of SN~1987A is shown for comparison.}
    \label{fig:80K_SED}
\end{figure*}

\begin{figure*}
    \includegraphics[width=0.48\textwidth]{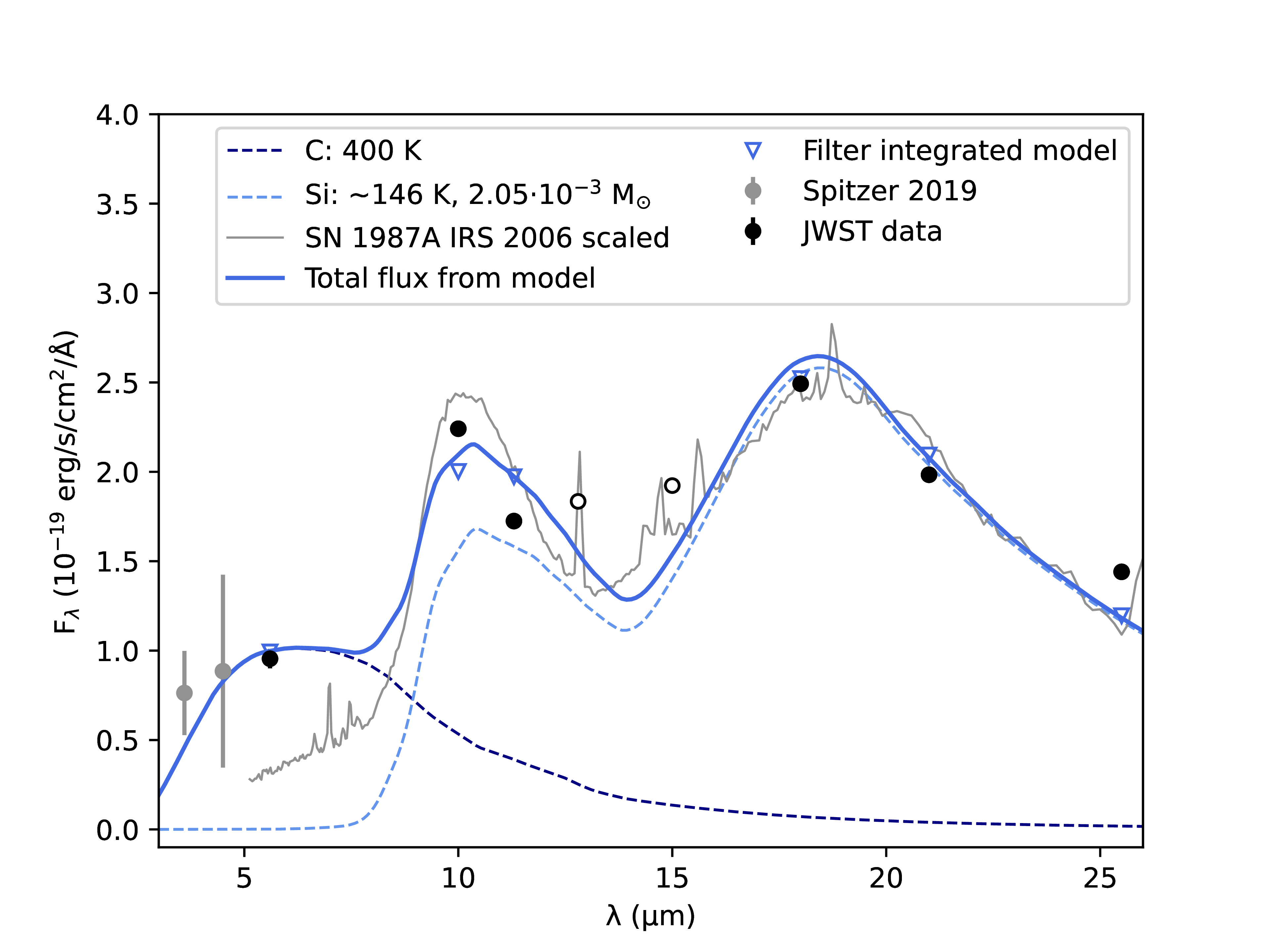}
    \includegraphics[width=0.48\textwidth]{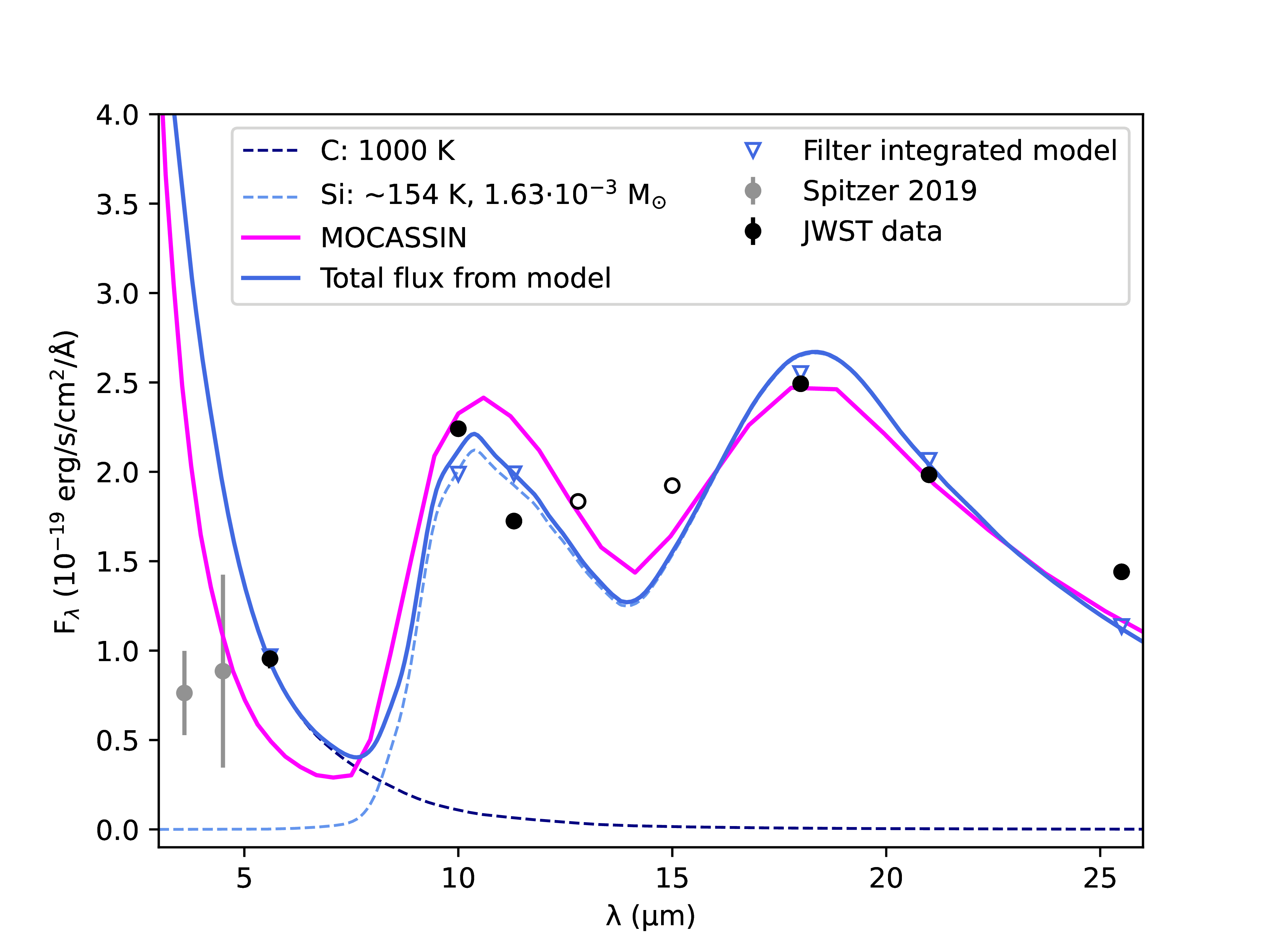}
    \caption{Best-fit two-component dust model fitted to the mid-IR SED of the SN with assumed cool silicate and a hot carbon component; we used a fixed value for $T_\rmn{hot} = 400$~K (on the left) and $T_\rmn{hot} = 1000$~K (on the right). We performed the fitting excluding the 12.8 and 15.0~$\mu$m data points because of assumed line emission (see text for details). We also present reanalysed previous {\it Spitzer}/IRAC 3.6 and 4.5~$\mu$m data for SN~1980K obtained in 2019 --- $\sim 39$~yr after explosion --- as comparison (these points, owing to their large uncertainties, were not used during the fitting process; see details in the text). We marked the total flux from the analytical model with a blue line.
    Black dots are the {\it JWST} data (empty circles represent the excluded data points), blue triangles are the filter-integrated model fluxes, the grey line represents the late-time scaled {\it Spitzer}/IRS spectrum of SN~1987A \citep{bouchet06}, and the purple line represents an output of our \textsc{mocassin} modeling (using the dust mass derived from our analytical SED modeling as an input parameter).}
    \label{fig:80K_SED_final}
\end{figure*}

During the analytical modeling of the {\it JWST}/MIRI SED of SN~1980K, we assume that the source of the observed late-time mid-IR flux is the thermal radiation from local dust grains. 
We basically followed the method described by \cite{Shahbandeh2023}, adopting the widely-used model of \cite{hildebrand83}, which assumes only the thermal emission of an optically thin dusty shell at a single equilibrium temperature $T_d$ and distance $D$, with a dust mass of $M_d$ and particle radius of $a$. 
The observed flux of the dust can be written as

\begin{equation}
    F_{\lambda} = \frac{M_d B_{\lambda}(T_d) \kappa(a)}{D^2}\, ,
    \label{eq:flux}
\end{equation}

\noindent where $B_{\lambda}(T_d)$ is the Planck function and $\kappa(a)$ is the dust mass absorption coefficient as a function of the dust grain radius defined by

\begin{equation}
    \kappa(a) = \left( \frac{3}{4 \pi \rho a^3} \right) \pi a^2 Q_v(a),
    \label{eq:kappa}
\end{equation}

\noindent where $Q_v(a)$ is the emission efficiency and $\rho$ \citep[for silicates 3.3~g~cm$^{-3}$][]{sarangi22} is the bulk density. 
Note that, while \cite{Shahbandeh2023} adopted a more general formalism of \citet{Dwek_2019} allowing for the presence of optically thick dust, we focused only on the optically thin case because of (i) the decades-long age of the SN remnant, and (ii) the likely presence of silicate emission at $\sim 10$~$\mu$m would be suppressed by optically thick dust.

During the modeling process, we calculated filter-integrated fluxes using Eq. \ref{eq:flux} convolved with the {\it JWST}/MIRI filters transmission profile and fit them to the {\it JWST} data. As described in Sec. \ref{sec:obs_spitzer}, we also measured {\it Spitzer} fluxes obtained between 2004 and 2019. 
The last {\it Spitzer} data, taking into account the age of the SN, are roughly contemporaneous with the {\it JWST} measurements; however, owing to their large uncertainties, we did not use them when fitting the data.

Because of the lack of any mid-IR spectra, there is a limitation on finding the true chemical composition of the assumed dust in SN~1980K. 
Based on the very recent models of the evolution of SN ejecta dust \citep{sarangi22}, as well as on previous studies of well-studied SNe, such as SN~1987A \citep{bouchet06,dwek10,arendt14} or SN~1993J \citep{zsiros22}, either silicates or carbonaceous dust can be present in a CCSN --- or, in its environment --- a few decades after explosion. 
Thus, we tested both cases in this study.
Just as in \citet{Shahbandeh2023}, the absorption and emission properties for silicate and amorphous carbon grains are obtained from \cite{Draine_2007} and \cite{Zubko_2004}, respectively (see \cite{sarangi22} for the values of absorption coefficients $\kappa$).
\cite{fox10} and \citealp{sarangi22} showed that assuming small ($\lesssim 0.1$~$\mu$m) grains does not result in significantly different models and has a small impact on the inferred dust mass and temperature. 
We also tested models with smaller grains in the case of SN~1980K and found negligible differences in the model parameters. 
Therefore, we applied a 0.1~$\mu$m grain size throughout our analysis \citep[as also applied by, e.g.,][]{szalai19,zsiros22}. 

First, we found that the full eight-point {\it JWST}/MIRI SED of SN~1980K cannot be properly fitted with either one- or two-component dust models, primarily because of the flux excess found at the 5.6~$\mu$m data point.
Thus, in the following approach, we excluded the 5.6~$\mu$m data point --- this can be explained with another, hot component, see in Sections \ref{sec:anal_analysis} and \ref{sec:anal_numerical} --- and fit one-component silicate models to the rest of the seven MIRI fluxes.
This step seemed more promising; however, at some wavelengths, model fluxes still do not match sufficiently well with observed ones (see left panel of Fig. \ref{fig:80K_SED}). 
Including 2--3 more dust components may result in a proper fit of the SED, but assuming more components with slightly different temperatures is difficult to explain and requires too many free parameters. 

Alternatively, since we are assuming an optically thin case, we checked the possibility of a line-emission contribution to the mid-IR SED.  
As mentioned above, there are no available mid-IR spectra of SN~1980K or any other SNe decades after explosion except {\it{Spitzer}}/IRS spectra of SN~1987A and the Type IIb SN~1993J. 
While the only late-time IRS spectrum of SN~1993J \citep{zsiros22} is quite noisy and covers only the 5--14~$\mu$m range, we used an IRS spectrum of SN~1987A dominated by emission of circumstellar dust located in the equatorial ring (ER) \citep[obtained in 2006, at an age of $\sim 19$~yr;][]{bouchet06} for comparison. 
After correcting for distances \citep[adopting $D=51.4$~kpc for SN~1987A from][]{bouchet06} and rescaling the temperature of the spectrum of SN~1987A from 180~K to 160~K (by multiplying the flux of SN~1987A by the ratio of the Planck function at 160~K to that at 180~K), we found a reasonably good match between the two datasets -- except for the quite different flux levels (due to the very different dust masses; see below).

Following the spectrum of SN~1987A, the flux excess around 12.8, 15.0, and 25.5~$\mu$m seen in the {\it JWST}/MIRI data of SN~1980K could be well explained by the line emission of [Ne~{\sc ii}] 12.81 $\mu$m, [Ne~{\sc iii}] 15.56 $\mu$m, and a complex contribution of [Ne~{\sc v}], [F~{\sc iv}], [O~{\sc iv}], and [Fe~{\sc ii}] around 25.5~$\mu$m, based on \cite{bouchet06}.
Excluding these three points from the fitting (still using a one-component Si-dust model), we obtain a dust temperature and mass similar to the previous one, while achieving a better match for the remaining wavelengths (see the right panel in Fig. \ref{fig:80K_SED}).

Lastly, we added a fixed temperature hot amorphous carbon component to fit the 5.6~$\mu$m data. 
\citet{dwek10} found that part of the late-time mid-IR spectra of SN~1987A below $\sim 8$~$\mu$m can be only fit with a secondary, significantly higher-temperature ($T \approx 350$~K) dust component arising from the ER. 
Similar cases were found  by \cite{Shahbandeh2023}, who used a $T=1000$~K hot carbon dust component in order to fit the 5.6~$\mu$m MIRI fluxes of both SN~2004et and SN~2017eaw. 
Thus, we performed fits with two different hot-component temperatures ($T_\rmn{hot} = 400$~K and 1000~K).

We calculated the confidence intervals for the parameters using \textit{lmfit.conf\_interval}\footnote{\href{https://lmfit.github.io/lmfit-py/confidence.html}{https://lmfit.github.io/lmfit-py/confidence.html}}. Moreover, applying two-component models implies that estimating the uncertainty from the covariance matrix is not appropriate \citep[also see, e.g.,][]{Shahbandeh2023}.
Our best-fit models are presented with a $\pm1\sigma$ confidence interval in Fig. \ref{fig:80K_SED_final} (see Table \ref{tab:dustpar}).
The primary (silicate) component has a mass of $M_\rmn{cold} \approx 2.0 \times 10^{-3}$~M$_{\odot}$ and a temperature of $T_\rmn{cold} \approx 150$~K.

The fixed secondary component ( $T_\rmn{hot} = 400$~K and 1000~K in the two different models) helps provide a better fit at 5.6~$\mu$m. Using either of the two hot components contributes only slightly to the longer-wavelength part of the mid-IR SED and results in the same values of the cold-component parameters within uncertainties.
Based on {\it JWST} data alone, we cannot conclusively determine the temperature of the warm component. If we consider the additional {\it Spitzer} 3.6~$\mu$m and 4.5~$\mu$m fluxes (which are {\it not} involved into the fit), $T_\rmn{hot} = 400$~K seems to be a more plausible choice. Nevertheless, both the high uncertainties of the {\it Spitzer} data and our numerical models (see later) allow the presence of a hot component with an even higher temperature.

We should also point out further limitations of the applied dust model. 
This model is based on the assumption that the observed flux arises from an optically thin dust shell, which may not be the case in general.
This assumption may lead to an underestimation of the total dust mass \citep{priestley20}. 
Moreover, \cite{Dwek_2019} discussed that a large amount of dust could fit the data in the case of an increasing optical depth.
However, to check the self-consistency of our models \citep[see also][]{fox10, zsiros22}, we calculated the optical depth as follows \citep{lucy89}:

\begin{equation}
    \tau = \kappa_{\textrm{average}} \frac{M_{\textrm{dust}}}{4 \pi r^2}\, ,
    \label{eq:tau}
\end{equation}

\noindent where we used $\kappa_{\textrm{average}}=750$~cm$^2$~g$^{-1}$ estimated for 0.1~$\mu$m silicate grains \citep[based on grain properties published by][]{Draine_2007,sarangi22}. 
In the case of optically thin dust, the blackbody radius sets the minimum shell size of an observed dust component.
Computing this blackbody radius, $R_\rmn{BB}$, for the primary (silicate) dust component, we get $2.1 \times 10^{16}$~cm. 
Using this value and the dust mass from our best model described above, we find $\tau \approx 0.27$; thus, we obtained an optically thin dust shell, which is in accordance with our initial assumption.

However, analytical model components describe only the total dust mass at an average dust temperature without reference to its relative geometry.
Even if we assume that the primary (cool) dust component is located in the ejecta, its temperature ($T_\rmn{cold} \approx 150$~K) is too high not to involve any extra heating mechanisms. 
As described in detail by e.g. \cite{sarangi22}, dust grains in the ejecta are heated radiatively from the decay of radioactive isotopes, and, much less dominantly, through collisions with the ambient gas. Decades after explosion, the radioactive heating of the SN is dominated by the decay of $^{44}$Ti (and, less dominantly, of $^{60}$Co); for SN~1987A, a total luminosity of less than $10^3$~L$_{\odot}$ was calculated by (for example) \citet{fransson02} and \citet{Seitenzahl14}.
This value is generally used as an order-of-magnitude approximation for CCSNe at the transitional phase \citep[see, e.g.,][]{tanaka12,Shahbandeh2023}; thus, we also use it as an upper limit for the contribution of radioactive heating to the late-time luminosity of SN~1980K.
The total luminosity calculated from the observed mid-IR SED is an order of magnitude higher ($L_\rmn{IR} \approx 2.4 \times 10^{38}$~erg~s$^{-1}$, i.e. $6.3 \times 10^4$~L$_{\odot}$), thus, it is necessary to invoke additional heating sources --- such as back-scattered high-energy photons from the CSM interaction --- or assume that the dust we see in the mid-IR is located in the CSM and is heated collisionally/radiatively. 
We discuss further details in Section \ref{sec:disc}.

Since certain dust-forming and heating processes take place in different regions of (or, outside) the expanding remnant, the location of the dust plays an important role in identifying its origin. 
In order to draw conclusions on the location of the assumed dust grains, we computed numerical models with a radiative-transfer code and also investigated the critical radii of the SN together with the possible dust-forming and heating channels.

\begin{table}
	\centering
	\caption{Best-fit model parameters of two-component dust models in case of the two different values adopted for the hot component.$^a$} 
	\label{tab:dustpar}
	\begin{tabular}{lccr} 
		\hline
		$T_\rmn{cold}$ & $M_\rmn{cold}$ & $T_\rmn{hot}$ & $M_\rmn{hot}$\\
        (K) & ($10^{-3}~{\rm M}_{\odot}$) & (K) & ($10^{-6}~{\rm M}_{\odot}$)\\
        \hline
		  $146.0_{-6.7}^{+6.6}$ & $2.05_{-0.68}^{+0.50}$ & 400 & 6.6\\
        $153.6_{-6.1}^{+6.3}$ & $1.63_{-0.39}^{+0.49}$ & 1000 & 0.1\\
		\hline
	\end{tabular}
 
 $^a$We assumed a cool silicate component and an additional fixed-temperature hot amorphous carbon component, and used a 0.1~$\mu$m grain size, for all models. Best-fit values are presented with $\pm1\sigma$ ($\sim 68$\%) confidence interval as reported by \textit{lmfit.conf\_interval}. The temperature of the hot component was fixed throughout the fitting process.
\end{table}

\subsection{Numerical modeling of the mid-IR SED}\label{sec:anal_numerical}

For a consistency check of our analytical models and also to determine the possible geometry of the dust, we applied the \textsc{mocassin}\footnote{ \href{https://mocassin.nebulousresearch.org/}{https://mocassin.nebulousresearch.org/}} \citep[MOnte CArlo SimulationS of Ionized Nebulae,][]{ercolano03,ercolano05,ercolano08} three-dimensional (3D) Monte Carlo photoionization and dust radiative-transfer code.
The code was originally developed to describe photoionization regions, but has been also applied in several cases for modeling dusty SNe \citep[e.g.,][]{ercolano07,Wesson15,Wesson21}.

In \textsc{mocassin}, the radiation field is described with a discrete number of energy packets, and the mean intensity is calculated from the possible light-matter interactions of these energy packets. 
The code follows the absorption and re-emission on account of particles in the gas and the absorption, scattering, and re-emission owing to dust grains \citep{ercolano05}. 
Dust grains with different sizes and species can be included and treated separately throughout the modeling. 
Numerous dust geometrical configurations can be considered in the code, including spherically symmetric, shell-like, and torus-like structures. 
The code can also take into account smooth and clumpy dust density distributions and handle various clump parameters. 

Various models with \textsc{mocassin} were computed to find similar outputs to the observed mid-IR SED of SN~1980K. 
We examined a spherical shell of dust, choosing a central illuminating source with a simple blackbody continuum, following other \textsc{mocassin}-based SN studies \citep[e.g.,][]{Ansari_2022}. 
(As described above, radioactive decay cannot heat the dust to the temperatures that we observe, implying that there is no special need to choose a diffuse heating source.) 
In this framework, the luminosity, the temperature of the illuminating source, and the size of the dust shell are not independent parameters. 
However, we found that the shape of the IR SED is mainly insensitive to the temperature of the central source; similar results were presented in the case of SN~1987A on day 8515 after explosion \citep{Wesson15}. 
Moreover, we were able to fix the luminosity of the central source close to the observed mid-IR luminosity of the SN ($L_\rmn{IR} \approx 3 \times 10^{38}$~erg~s$^{-1}$), while its temperature was finally set to 4800~K (giving a mid-IR SED with a shape being similar to the observed one).

We assumed both silicate and amorphous carbon composition with a grain radius of 0.1~$\mu$m, and a particle density distribution proportional to $r^{-2}$. 
The same extinction, absorption, and scattering coefficients used in the case of analytical models \citep{sarangi22} were applied. 
We fixed the dust mass at $1.8 \times 10^{-3}~{\rm M}_{\odot}$ as an average value for the amount of cold dust originating from our final analytical models.

After doing these steps, we were allowed to compute models in a large range of potential shell sizes, varying the values of the outer radius and the ratio of inner and outer radii. Since we do not know the true nature of the heating source of the dust grains in SN~1980K, we cannot be sure how realistic these sizes are; nevertheless, we were able to test, at a given total luminosity and dust mass, what (if any) kind of dust geometry provides outputs comparable to the observed mid-IR SEDs.

Similarly to the results of our analytical modeling (see in Fig. \ref{fig:80K_SED_final}), \textsc{mocassin} models do not match the {\it JWST} data in the case of carbon composition dust. 
Thus, numerical models strengthen the conclusion that the dust in the environment of SN~1980K is presumably dominated by silicates. 
We found that the SN can be described with an outer radius of $R_\rmn{out} \approx 1.5\times10^{17}$~cm, and a ratio of the inner and outer radii of $R_\rmn{in}/ R_\rmn{out} \approx 0.08$, implying that the presumed dust shell can be quite thick.
This configuration results in a very low optical depth, strengthening the optically thin assumption we applied during our analytical modeling.

We also found that, at shorter wavelengths, \textsc{mocassin} models show a large flux excess, regardless of the values of the assumed parameters. 
This aligns well with the results of our analytical models.
In the right panel of Fig. \ref{fig:80K_SED_final}, we compare a silicate-based \textsc{mocassin} model with our best-fit two-component analytical model containing a 1000~K hot component.

\subsection{Modeling optical line-profile asymmetries}\label{sec:anal_optspec}

Another essential part of this study is the analysis of the late-time optical spectrum of SN~1980K (see Sec. \ref{sec:obs_optspec}), in order to find further clues about local dust and to compare the results with those of the mid-IR SED analysis.
It is a general assumption that red-blue asymmetries in optical and near-IR emission-line profiles may be associated with the presence of newly-formed dust in the (inner) ejecta of CCSNe \citep{lucy89,milisavljevic12,bevan16,bevan17,niculescu-duvaz22}. 
In this picture, the presence of SN dust results in an extended red wing and/or a blueshifted peak due to scattering and absorption effects. 
Specifically, dust in the ejecta affects light only from the receding part of the ejecta, causing the red wing of the line profile to be suppressed. 
Crucially, CSM dust does not do this, being exterior to the expanding ejecta; it absorbs light from both the approaching and receding gas.
As an alternative explanation, red-blue asymmetries of the line profiles may originate from intrinsic asymmetries in the SN explosion; however, if this were true, we would see a statistically homogeneous sample of redshifted and blueshifted emission-line peaks in SN spectra --- and this is not the case \citep{niculescu-duvaz22}.

\cite{bevan16} present a code called \textsc{damocles} (Dust Affected Models Of Characteristic Line Emission in Supernovae) to model the effect of dust on line profiles of CCSNe. 
With \textsc{damocles}, it is possible to investigate the effect of dust internal to SNe (hence located in the shocked and/or unshocked ejecta) on the line shapes and determine the dust properties. 
\textsc{damocles} is based on a Monte Carlo approach and can handle various grain species and sizes with smooth and clumpy dust density profiles. 
The recent Keck spectrum provides a great opportunity to draw conclusions about the dust parameters from the optical line profiles and also a comparison of our findings with those of previous studies \citep{milisavljevic12,bevan17,niculescu-duvaz22}. 

\begin{figure}
        \includegraphics[width=0.48\textwidth]{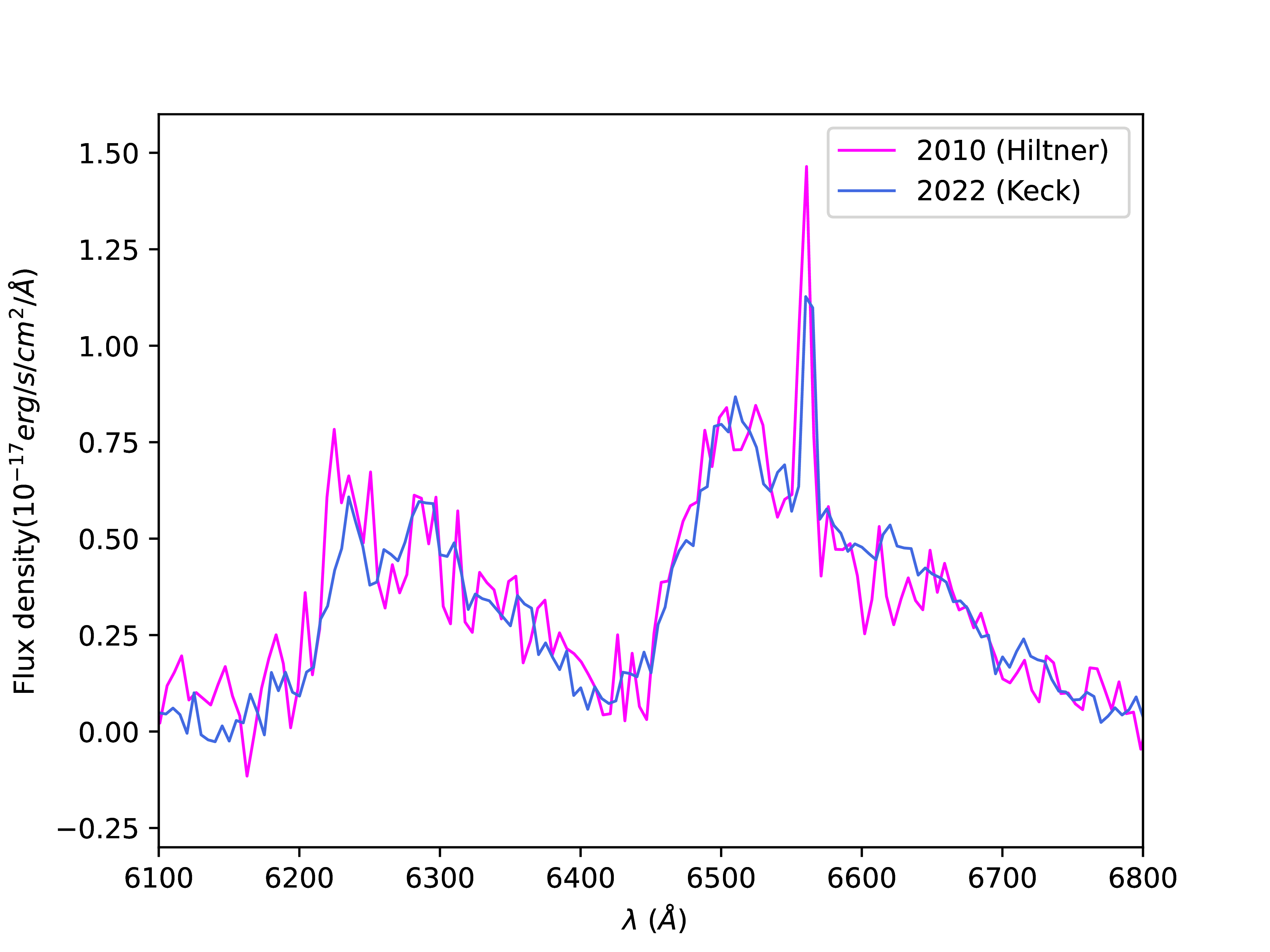}
    \caption{[O~{\sc i}] and H$\alpha$ line profiles in the 5 \AA -binned spectrum of SN~1980K obtained with Keck/LRIS on 2022 Nov. 19 (blue) compared to the spectrum from 2010 \citep[with purple line, obtained with the 2.4m Hiltner Telescope at MDM Observatoy on Kitt Peak, Arizona; see][]{milisavljevic12,bevan17}.}
    \label{fig:80K_sp}
\end{figure}

The code does not use an entirely self-consistent procedure; instead, it assumes that all absorbed energy packets may be re-emitted outside of the explored wavelength range. 
Moreover, the optical extinction due to dust is considered to be temperature independent and the total energy is not conserved in the examined wavelength range, the code retains the signature of the normalised line profiles \citep{bevan16}. 

The ejecta are assumed to be in a 3D Cartesian grid with a particular composition and density distribution. 
The gas functions as an initial emission source, and dust properties are considered throughout the analysis \citep{bevan16}. 
We coupled the gas and dust and did not included electron scattering. 
If we assume a constant SN induced mass-flow rate, the velocity and the dust density exponents become dependent upon each other \citep{bevan16}.

Following the previous studies, we investigate the H$\alpha$ and [O~{\sc i}] $\lambda\lambda$6300, 6363 line profiles for modeling with \textsc{damocles}. 
Our modeling approaches basically follow the methods described by \cite{bevan16} \citep[also applied by][]{bevan17, niculescu-duvaz22}. 
Before investigating the models, we compared the new Keck spectrum of SN~1980K to the ones obtained in 2010 \citep{milisavljevic12,bevan17} and during 2016--2018 \citep{niculescu-duvaz22}. 
We found that our recently obtained spectrum is very similar to previous ones (Fig. \ref{fig:80K_sp}); however, there were some absolute flux calibration and signal-to-noise-ratio issues between the 2016--2018 average spectrum presented by \citet{niculescu-duvaz22} and our one.
Thus, we used only the 2010 spectrum and the work of \citet{bevan17} as a reference during the next steps of \textsc{damocles} modeling. 
Nevertheless, the dust masses presented by \citet{niculescu-duvaz22} (0.2 and 0.6~\msolar\ in the case of smooth and clumpy line-profile models for both lines, respectively) basically agree with our results (Table \ref{tab:lines_pars}).
 
Note that the dust masses depend on several model parameters, especially in the case of clumpy models. Furthermore, the [O~{\sc i}] $\lambda\lambda$6300, 6363 line already has a complex profile. Thus, we focused on retrieving the minimal dust masses required to match the shape of the line profile with the observed spectra.
Comparing the H$\alpha$ and [O~{\sc i}] $\lambda\lambda$6300, 6363 line profiles of the two spectra in detail, we only found a small change in the velocities of [O~{\sc i}] lines (5000~km~s$^{-1}$ instead of 5500~km~s$^{-1}$), while for the H$\alpha$ line, we defined the same outer expansion velocity (5500~km~s$^{-1}$) as \cite{bevan17}.
We assumed free expansion and calculated the outer radius of the emitting region from the velocity profile. 
Since both lines are optically thin recombination lines, we considered the emissivity distribution proportional to the square of the local gas density \citep{bevan16}. 

Both amorphous carbon and silicate compositions were tested with smooth and clumpy density distributions, exploring the parameter space manually. 
Dust grains are more likely to accumulate in clumps \citep[e.g.][]{Wesson15}, however, in this case, radiation is exposed to less scattering and to less absorption. 
Thus, clumpy \textsc{Damocles} models yield less of the scattering wing and attenuate the red wing less than expected for a smooth model and require a factor of 1.5--3$\times$ larger dust masses and also a small adjustment to the grain sizes \citep{niculescu-duvaz22}. 
We applied grains with 0.1~$\mu$m radius, according to both our analytical and \textsc{mocassin} models, and previous \textsc{Damocles} models \citep{bevan17,niculescu-duvaz22}. 
Contrary to the radiative-transfer modelling of the IR SED --- where greater grain sizes typically lead to lower dust masses to fit the IR regime \citep[see][]{Wesson15} --- in the case of optical line-profile analysis, larger dust masses are needed to fit the same amount of absorption in the case of larger grain sizes \citep{bevan16}.

First, we tested line-profile models adopted from \cite{bevan17} taken 10,964 days post-explosion and evolved to 15,388 days post-explosion to describe the Keck spectrum. 
We found that in the case of smooth and clumpy density distributions, a higher dust mass is required to fit both the H$\alpha$ and [O~{\sc i}] $\lambda\lambda$6300, 6363 line profiles of the Keck spectrum, while other parameters needed fine-tuning. 
\cite{bevan17} reported 0.1~M$_{\odot}$ and 0.12~M$_{\odot}$ of dust in the case of H$\alpha$ line-profile models, and 0.2~M$_{\odot}$ and 0.3~M$_{\odot}$ dust in the case of [O~{\sc i}] $\lambda\lambda$6300, 6363 line-profile models, for smooth and clumpy dust density distributions, respectively.
To find the best model, we also varied all the examined parameters. 
First, we identified the outer expansion velocity that describes the width of the broadened line. 
Then, we set the corresponding density profile index (defined by $\rho \propto r^{-\beta}$) and dust mass based on the steepness of the red wing.

\begin{figure*}
        \includegraphics[width=0.45\textwidth]{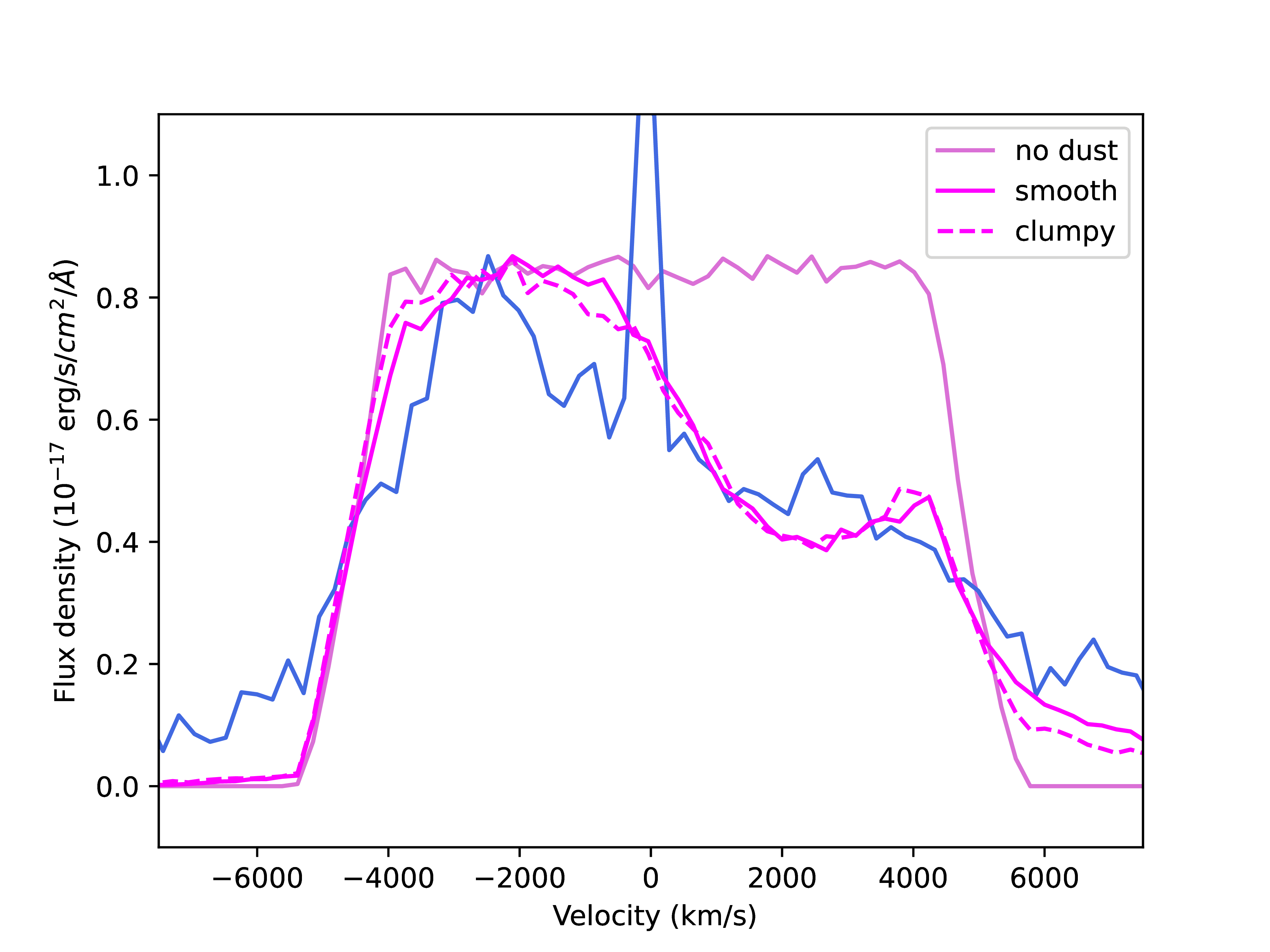}
        \includegraphics[width=0.45\textwidth]{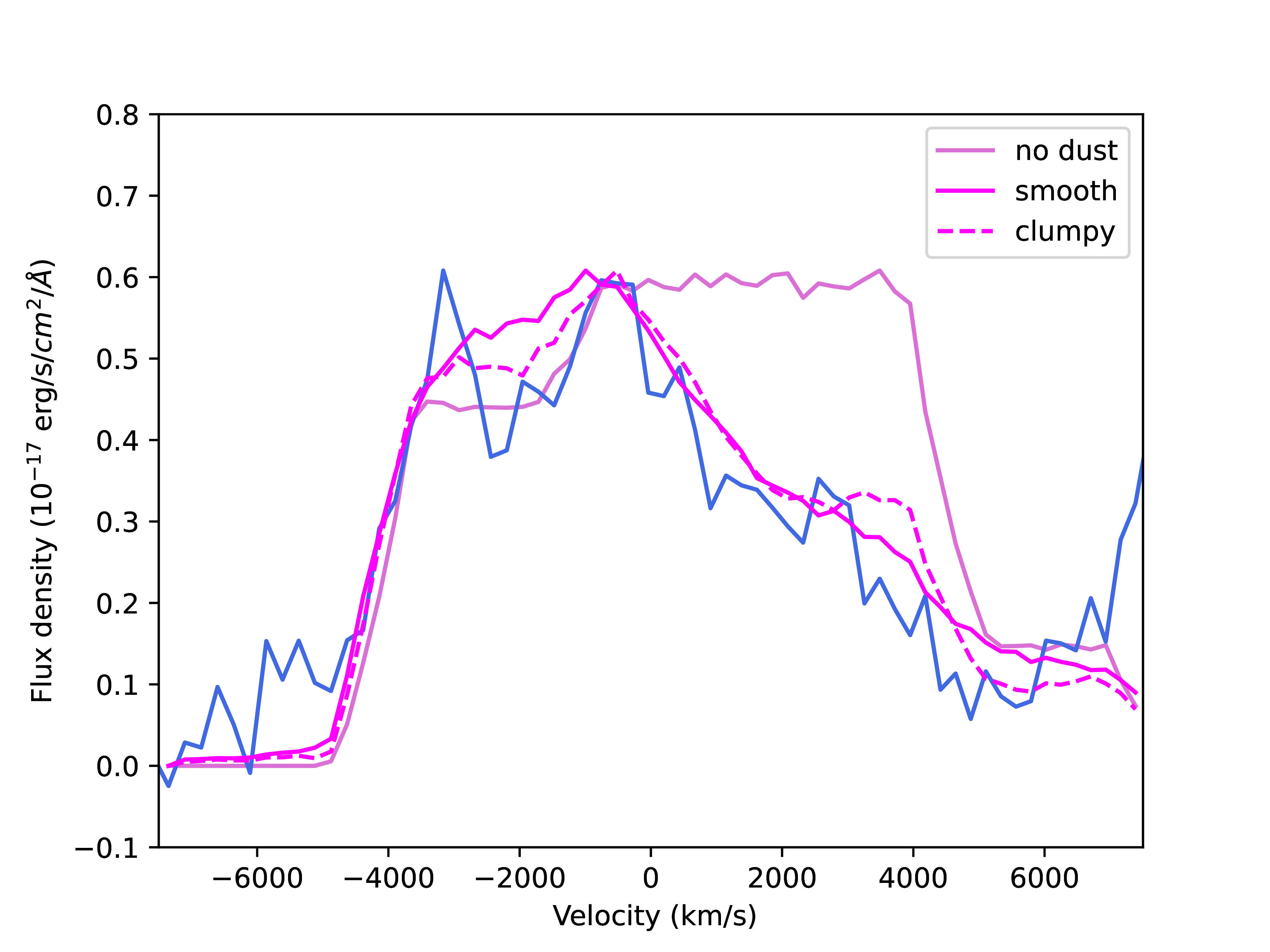}
    \caption{The best smooth and clumpy (continuous and dashed purple lines, respectively) silicate dust models along with the dust-free (pink line) model for the H$\alpha$ and [O~{\sc i}] $\lambda\lambda$6300, 6363 (5 \AA -binned) line profiles in SN~1980K (blue line).}
    \label{fig:80K_damocles}
\end{figure*}

Our line-profile models align very well with those presented by \cite{bevan17}.
Our best line-profile models are shown in Fig. \ref{fig:80K_damocles}.
In the case of the best H$\alpha$ line-profile models, we found 0.24~M$_{\odot}$ and 0.38~M$_{\odot}$ of dust for smooth and clumpy density distributions, respectively. 
In the case of the best [O~{\sc i}] $\lambda\lambda$6300, 6363 line-profile models, we found about a factor of two larger dust masses: 0.37~M$_{\odot}$ and 0.58~M$_{\odot}$ dust for smooth and clumpy density distributions, respectively. 
While the overall H$\alpha$ line profile remains unchanged, the [O~{\sc i}] $\lambda\lambda$6300, 6363 line profiles show definite evolution compared to the $\sim 30$~yr spectrum. 
We also found that [O~{\sc i}] $\lambda\lambda$6300, 6363 line-profile models result in lower velocities than the H$\alpha$ models, so the line-forming regions are more likely located in a more inward part of the ejecta for the [O~{\sc i}] lines.

\begin{table*}
	\centering
	\caption{Model parameters used for \textsc{damocles} line-profile models.$^a$}
	\label{tab:lines_pars}
	\begin{tabular}{lccccccr} 
		\hline
		Line & Clumpy & $v_\rmn{max}$ & $R_\rmn{in}/$ & $R_\rmn{out}$ & $R_\rmn{in}$ & $\beta$ & $M_\rmn{dust}$\\
         &  & (km~s$^{-1}$) & $R_\rmn{out}$ & ($10^{17}$~cm) & ($10^{17}$~cm) &  & (M$_{\odot}$)\\
        \hline
		H$\alpha$ & no & 5500 & 0.75 & 7.3 & 5.5 & 2.0 & 0.24\\
        H$\alpha$ & yes & 5500 & 0.75 & 7.3 & 5.5 & 2.0 & 0.38\\
        $[\rm{O I}]$ & no & 5000 & 0.75 & 6.6 & 5.0 & 2.0 & 0.37\\
        $[\rm{O I}]$ & yes & 5000 & 0.75 & 6.6 & 5.0 & 2.0 & 0.58\\
		\hline
	\end{tabular}

 $^a$We assumed 100\% silicate dust for all models with 0.1~$\mu$m grain size. A 3.1 doublet intensity ratio was adopted for the [O~{\sc i}] $\lambda\lambda$6300, 6363 lines. In the case of clumpy models, we used filling factor $f=0.1$, and the model grids were divided into 40 steps in each axis.
\end{table*}

\cite{bevan17} found that the H$\alpha$ and [O~{\sc i}] $\lambda\lambda$6300, 6363 line profiles are widely similar, but the oxygen models definitely require larger dust masses. 
Thus, they suggest that dust-forming regions may be focused in the vicinity of oxygen-rich zones \citep[the same as found in the case of SN1987A;][]{bevan16}. 
In this way, a more complex model, assuming diffuse hydrogen and clumped oxygen with a decoupled distribution of dust and gas, possibly could resolve the discrepancy in the dust masses \citep{bevan17}. 
However, these points are beyond the scope of this work. 

\section{Discussion \& Conclusions}\label{sec:disc}

In this section, we summarise all of our conclusions regarding the amount, origin, and geometry of the dust located in (or, around) SN~1980K. 

Determining the possible location of the presumed dust shell(s) is a vital step to revealing the potential origin of the dust. 
Newly-formed grains could be located in the inner (unshocked) ejecta and in the CDS between the outer (shocked) ejecta and shocked CSM. 
On the other hand, more distant pre-existing dust can be heated collisionally (i.e., via direct interaction with the forward shock) or radiatively (by the peak SN luminosity --- however this option does not seem significant more than 40 yr post-explosion --- 
and/or by energetic photons arising from CSM interaction). 

\subsection{Collisionally heated CSM dust scenario}\label{subsec:col_disc}

The similarity between the late-time mid-IR SED of SN~1980K and the {\it Spitzer}/IRS spectrum of SN~1987A (described in detail in Sec. \ref{sec:anal_analysis}) leads us to examine the possibility of collisionally heated CSM dust as the first step.
In a sequence of papers, \citet{dwek10}, \citet{arendt16}, and \citet{arendt20} found that the mid-IR ($\sim 5$--30~$\mu$m) component of the radiation of SN~1987A probably originates from the collisional interaction of the SN shock and some pre-existing matter located in an ER around the SN. 
This portion of the SED of SN~1987A was fitted by the cited authors with a dust component showing a very similar temperature ($T \approx 180$~K) to what we found in the case of SN~1980K; however, the mass of this dust component is $\sim 2$ orders of magnitudes larger in the case of SN~1980K. 
It is also in agreement with the evolution of mid-IR fluxes: since, in the case of SN~1987A, 5--30~$\mu$m fluxes rise and fall between $\sim 15$~yr and $\sim 30$~yr post-explosion (i.e., as the SN’s blast wave has run into and through the pre-existing circumstellar ER), the level of mid-IR emission is still high in the case of SN~1980K more than 40~yr post-explosion (and, based on the detailed analysis of both the {\it Spitzer} dataset and the optical spectra, degree of interaction could be more or less constant in the last $\sim 15$~yr, see above). 

So, if we accept the scenario of collisionally heated pre-existing dust (at least in the case of the $\sim 10^{-3}$~\msolar\ dust we see with {\it JWST}), then we potentially see a similar case to SN~1987A's ER but at a larger scale.

In order to examine whether pre-existing dust grains can be really heated by the post-shocked gas due to the collision of the ejecta and the CSM in the case of SN~1980K, we follow the method presented by \cite{fox10} and also applied by others\citep[e.g.,][]{fox11,tinyanont16,zsiros22} for estimating the mass of dust processed by the forward shock of the SN. 
Pre-existing grains are expected in regions farther from the SN, characterised by the dust-free cavity and the outer dust shell. 
Following a SN explosion, its peak luminosity destroys innermost pre-existing dust grains, thereby creating a dust-free cavity around the SN, and heats the remaining dust grains located between the created dust-free cavity and the outer radius of the external dust shell. 
The vaporisation radius ($R_{\rmn{evap}}$) can be calculated from the peak UV-visual luminosity of the SN \citep{dwek85}. 
Unfortunately, there are no applicable estimations for $L_{\rmn{peak}}$ in early-time data, so we cannot draw conclusions on the size of the dust-free cavity. 
Based on previous studies \citep[see, e.g.,][]{fox10,zsiros22}, this value is likely around a few times 10$^{16}$~cm, similar to the $R_\rmn{BB}$ value we determined above for SN~1980K. 
However, the size of the evaporation radius does not have significance, since the shock passed it a long time ago.

\citet{fox10} assume that the hot, post-shocked gas heats the dust shell and the total gas mass can be determined by the volume of the emitting shell applying the equations characterizing grain sputtering.
Considering a dust-to-gas mass ratio of 0.01, the mass of dust is

\begin{equation}
\label{eq:one}
M_{\mathrm{d}}({\rm M}_{\odot}) \approx 0.0028 \left( \frac{v_{\mathrm{s}}}{15,000 \,\mathrm{km} \,\mathrm{s}^{-1}} \right)^3 \left( \frac{t}{\mathrm{yr}} \right)^2 \left( \frac{a}{\mathrm{\mu m}}\right)\, ,
\end{equation}

\noindent where $v_\mathrm{s}$ is the shock velocity (assumed to be constant), $t$ is the time post explosion, and $a$ is the dust grain size.
We used $v_\mathrm{s} = 5000$~km~s$^{-1}$ and 15,000~km~s$^{-1}$ (respectively) for the lower and upper limits of the shock velocity \citep[taking the lower limit from our optical line analysis and the upper limit from][]{fox10}. 
We applied 0.005~$\mu$m and 0.1~$\mu$m for the lower and upper limits, respectively. 
On the basis of these assumptions, we find that 42~yr after explosion, the dust mass that can be heated collisionally is in the range 0.001--0.5~M$_{\odot}$. 
The lower end is close to the mass we deduce from our best-fit analytic dust models.
Thus, these results seem to support (or, at least, not to exclude) the possibility of the collisional heating of pre-existing grains.

We also compared our predictions for the dust emission with the mass-loss-rate estimates for SN~1980K by \cite{RizzoSmith23}. They present a mass-loss rate of $\sim 3.0 \times 10^{-7}$ M$_{\odot}$~yr$^{-1}$, which is considered to be a typical value for average CCSNe. 
However, the velocity from the H$\alpha$ line ($\sim 5500$~km~s$^{-1}$, obtained from the late-time optical spectra) suggest a considerable deceleration of the shock over the $\sim 40$~yr of evolution, which requires a higher mass-loss rate. Also, \cite{weiler92} report a larger mass-loss rate and therefore it cannot be excluded for the early phases.

Moreover, we make some additional comments on the hot (dust) component in the mid-IR SED of SN~1980K. 
As mentioned above, in the cases of both SNe~1980K and 1987A, we see at least two components in the 5--30~$\mu$m SED: one with a temperature of $\sim 150$~K (180~K) and one with $\sim 400$~K (or, with an even higher temperature in the case of SN~1980K; see Sec. \ref{sec:anal_analysis}). 
Based on studying contemporaneous IR and X-ray data of SN~1987A, \cite{dwek10} and \cite{arendt16} concluded that the colder (Si-dominated) component is probably collisionally heated by the hot circumstellar gas in the ER, while they did not find clear evidence for the origin of the hotter component (but excluded gas or the synchrotron emission mechanism as potential sources). 
In the case of SN~1980K, we do not have either direct images of an ER or late-time X-ray data; thus, the situation can be the same as in the case of SN~1987A, but it is also a possibility that the hotter (dust) component has a collisional heating source, while the colder component is coming from radiative heating of the pre-existing grains, or, from newly-formed dust in the ejecta (see below). 
Nevertheless, for a more detailed investigation, further data would be necessary --- especially in the range of $\sim 1$--5~$\mu$m to cover the assumed thermal radiation of the hot dust component.

\subsection{Radiatively heated CSM dust scenario}

If we assume the presence of CSM dust, another necessary step is to examine the possibility of radiative heating of the pre-existing dust grains around SN~1980K.
Considering a simple IR light-echo model of \cite{bode80} and \cite{dwek83}, as showed by \citet{fox10,fox11}, we can investigate the possibility of the presence of an IR echo.
If the dust is located in a shell between the vaporisation and echo radii (defined by $R_{\rm{echo}} = ct_{\rm{echo}}/2$, where $t_{\rm{echo}}$ is the duration of the echo \citep[see, e.g.,][]{fox10,fox11}, the late-time optical luminosity from the CSM interaction heats the grains. 
As mentioned above, unfortunately, we cannot determine the size of the dust-free cavity because of a lack of data. 
Nevertheless, the size of vaporization radius only has an importance when we assume that the pre-existing dust shell is heated by the SN peak luminosity. However, more than 40 yr post-explosion, the SN peak luminosity cannot be considered a significant heating source: assuming this type of heating and a radius of the dust shell being $\sim 10^{17}$~cm (estimated based on $R_\rmn{out}$ coming from our numerical SED modeling), it yields a few hundred-day-long light echo, which is not compatible with the current age of SN~1980K. 

On the other hand, seeing the obvious signs of long-term ongoing CSM interaction in the environment of SN~1980K, a CSM echo (i.e., reprocessing of the optical emission by the dust) may be a viable option. To examine this option in detail, we can follow again the method presented by \cite{fox10,fox11}.
First, we approximate the inner shell radius from the optical luminosity,

\begin{equation}
\label{eq:CSM_echo}
L_\mathrm{opt}= \frac{64}{3} \rho a R^2 \sigma T_\mathrm{SN} \frac{\int B_\mathrm{\nu}(T_\mathrm{d} \kappa(\nu)) d\nu}{\int B_\mathrm{\nu}(T_\mathrm{SN} Q_\mathrm{abs}(\nu)) d\nu}\, ,
\end{equation}

\noindent where $T_\mathrm{SN}$ is the effective SN blackbody temperature, $a$ is the grain size, $\rho$ is the dust bulk density, and $R$ is the echo radius. 
We used $T_\mathrm{SN} = 10,000$~K \citep[however, the results are relatively insensitive to the temperature; see][]{fox11}, $a = 0.1$~$\mu$m, and $\rho = 3.3$~g~cm$^{-3}$ (as previously for the silicate-based dust).

\begin{figure}
        \includegraphics[width=0.45\textwidth]{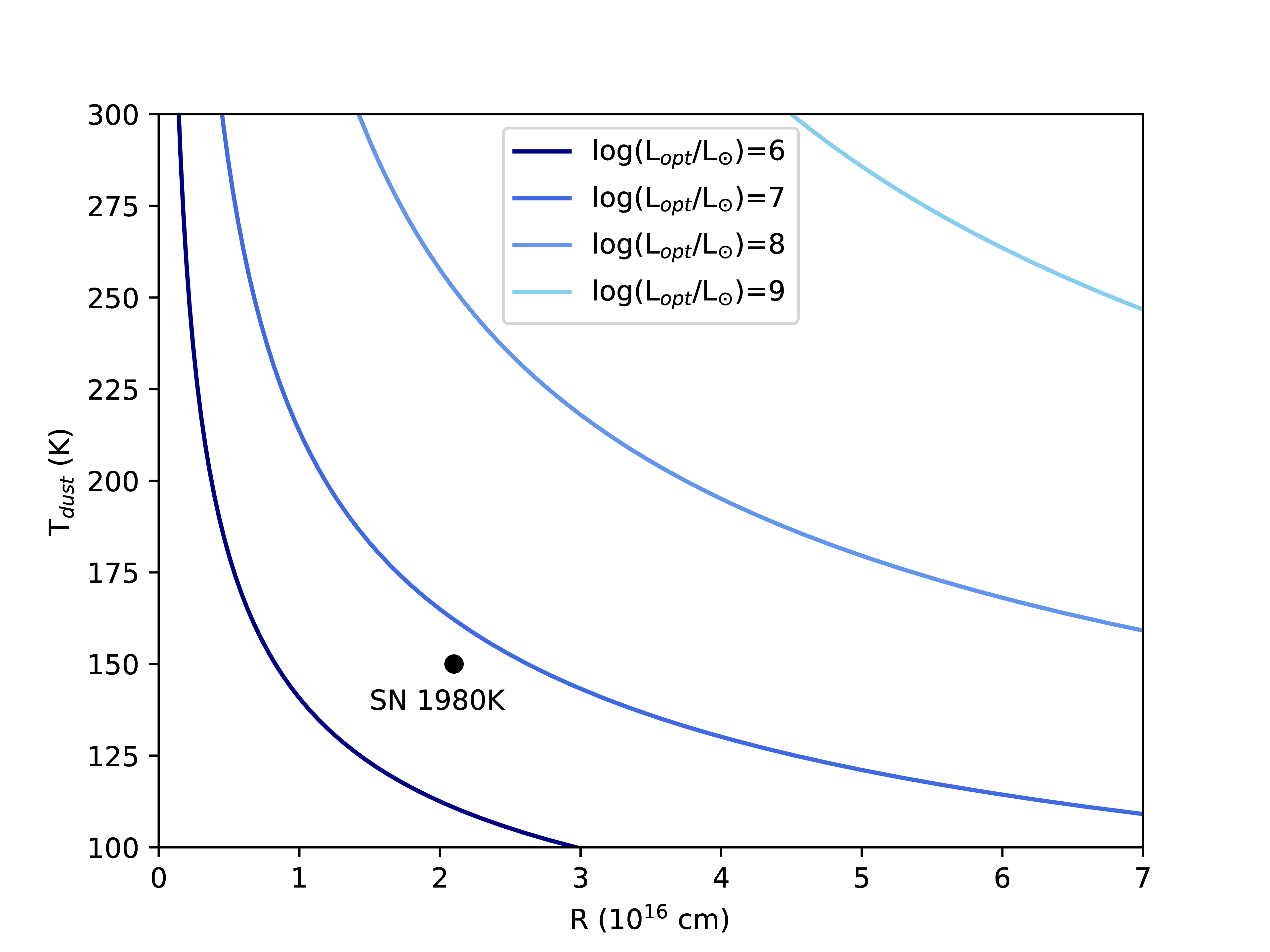}
    \caption{The expected average dust temperature as a function of the inner shell radius for several optical luminosities generated by CSM interaction. Blue lines represents the calculated model by Eq. \ref{eq:CSM_echo}.}
    \label{fig:echo}
\end{figure}

Following \cite{fox11}, to test the possibility of radiative heating by the CSM interaction, we compared the results of these calculations with the minimal (blackbody) dust radius ($R_\rmn{BB}$) we determined during our analytical SED modeling. 
Based on these calculations, a dust shell with a radius of $R_\rmn{BB} = 2.1 \times 10^{16}$~cm and a temperature of 150~K (heated by CSM interaction) requires a late-time optical luminosity of  $10^{6}~{\rm L}_{\odot} \le L_\rmn{opt} \le 10^{7}~{\rm L}_{\odot}$ (see Fig. \ref{fig:echo}).
Such large late-time optical luminosities are questionable in the case of SN~1980K since the integrated optical luminosity calculated from the Keck spectrum is only $\sim 1.3 \times 10^4~ {\rm L}_\odot$ (while the observed mid-IR luminosity, as described above, is $\sim 6.3 \times 10^{4}~{\rm L}_\odot$).

However, \cite{Dessart_2022} found that the shock power from the CSM interaction emerges primarily in the UV, and only a few percent of that emerges in the optical. 
Their models indicate that an optical luminosity of $\sim 10^4$~L$_{\odot}$ agrees with a UV luminosity of the CSM radiation of $\sim 10^5$~L$_{\odot}$.
While this value is still too low, continuous CSM interaction could be a contributing heating source for dust (being located even in the CSM or in the ejecta; see below). 

As a further possibility, according to optical and mid-IR ({\it{Spitzer}}/IRAC and MIPS) data taken between 2004 and 2008, \cite{sugerman12} suggest scattered and thermal light echoes off of more extended CSM in the environment of SN~1980K. 
They found that the SN shows a slow and monotonic fading which can be explained with echoes from a thin circumstellar shell 14--15 light year ($\sim 1.3$--1.4 $\times 10^{19}$~cm) from the progenitor containing $\leq 0.02~{\rm M}_{\odot}$ of carbon-rich dust. 
While the {\it JWST} data on SN~1980K do not contradict the possibility of such extended light echoes, we highlight here that the reanalysis of the full {\it Spitzer} dataset of the SN (see Fig. \ref{fig:80K_spitzer}) does not support the monotonic fading predicted by \cite{sugerman12}, thus weakening the hypothesis of scattered light echoes. 
Moreover, in another recent study, \cite{radica22} found no evidence of scattered light echoes in the vicinity of the SN according to spectral data of the SITELLE Imaging Fourier Transform Spectrometer on the Canada-France-Hawaii Telescope.

\subsection{Newly-formed ejecta dust scenario}

Based on the recent detailed calculations by \cite{sarangi22}, newly-formed dust should form at velocities $\lesssim 2500$~km~s$^{-1}$ in (Type IIP) CCSN ejecta. Using this velocity as an upper limit, we get a radius of a dust-forming region $R \lesssim 3.3 \times 10^{17}$~cm at 15,300 days. 
This value is (much) larger than the calculated blackbody radius ($R_\rmn{BB} \approx 2.1 \times 10^{16}$~cm), and agrees well with the outer shell radius from \textsc{mocassin} models assuming a spherical shell of dust heated by a central source ($R_\rmn{out} \approx 1.5 \times 10^{17}$~cm). 
Thus, it seems to support the assumption that all of the $\sim 2 \times 10^{-3}$~\msolar\ dust we see in the mid-IR can be placed within the ejecta and therefore can be purely newly-formed. 

However, the temperature of the dust imposes further constraints on our calculations, as we noted above. 
Taking a look again at the calculations by \cite{sarangi22}, the temperature of ejecta dust (assuming any kind of composition) should decrease below $\sim 100$~K up to 3000~days after explosion (the point at which the calculations end). 
Here, based on the mid-IR SED of SN~1980K, we see a primary (Si-dominated) dust component with a temperature of $\sim 150$~K more than 40~yr after explosion; this means that some additional source of heat is needed. 
Similar results were shown by \cite{Shahbandeh2023} for SNe~2004et and 2017eaw, where they mark the source of this extra energy as the interaction between the SN forward shock and the ambient CSM (emerging in the form of not just optical, but presumably also X-ray and UV photons even at these late epochs).
In the case of ejecta dust being heated by an external shock, the temperature and the total IR luminosity of dust depends on the relative radii of the dusty sphere and the shock. 
Simple back-of-the-envelope calculations show that silicate dust in the ejecta, confined within 2500~km~s$^{-1}$, will be at a temperature of $\sim 150$~K when heated by an external forward shock that has a velocity of $\sim 5000$~km~s$^{-1}$ (with reference to \citealt{fox11}; see \cite{Dessart_2022} for more details on shock luminosities). 
In these terms, the dust in SN ejecta is in agreement with the energetics required to produce the observed IR fluxes.

Beyond that, a further question remains open regarding the hot component we used to fit the 5.6~$\mu$m point of the mid-IR SED (can be $T \sim 400$~K, or even 1000~K, see Section \ref{sec:anal_analysis}). 
This component, if it truly originates from dust, is difficult to connect to any heating processes in the ejecta at such late times; however, as we noted in Section \ref{subsec:col_disc}, it may have a circumstellar origin (or, does not connect to the expanding SN ejecta at all).

Switching to the results of optical line-profile analysis (Sec. \ref{sec:anal_optspec}), these suggest an even larger amount ($\sim 0.24$--0.58~\msolar) of dust to be present in the ejecta. 
From \textsc{damocles} modeling, we do not obtain any information on the dust temperature; thus, it is required that {\it all of} the dust is in the ejecta and we see only the ``warm'' ($\sim 0.002$~\msolar, $T \approx 150$~K) component with {\it JWST} (the ``tip of the iceberg''), with much more colder dust only visible at longer wavelengths.

The location of the dust seems to be somewhat problematic here, since \textsc{damocles} models place the dust at (5.0--6.6) and (5.5--7.3) $\times 10^{17}$~cm --- from the analysis of the [O~{\sc i}] and H$\alpha$ lines, respectively (see Table \ref{tab:lines_pars}) --- assuming respective minimal velocities of 3750 and 4125~km~s$^{-1}$ \citep[which are too high compared to the calculations of][]{sarangi22}. 
At that point, it is worth examining the possibility of newly-formed dust in the CDS. 
Considering that the CDS is formed behind the reverse shock (in the Lagrangian reference frame), and that broad optical emission lines basically originate from the H-rich ejecta heated by the reverse shock, then both the front-side and back-side line emission would be absorbed by the dust; however, hydrodynamic mixing results in a clumpy CDS and enables some emission to pass through it \citep{milisavljevic12}.
Again, assuming a lower limit on the forward shock velocity to be $\sim 5000$~km~s$^{-1}$ (see above), a dense shell being present close to $\sim 4000$~km~s$^{-1}$ could be a reasonable estimate. However, the forward shock generally is nonradiative, but radiative shocks can occur in this region if the circumstellar gas is very clumpy \citep{chugai94,milisavljevic12}. 
Moreover, it also remains a matter of debate whether the CDS in such SNe can have enough mass of gas to host more than 0.1~\msolar\ of dust.

Nevertheless, as described in detail in Section \ref{sec:anal_optspec}, these values are the results of coupled gas-dust \textsc{damocles} models, and improved modeling via decoupling the two components might solve this problem.
We also note that the calculations of \cite{sarangi22} we use here pertain to Type IIP explosions. 
There are no such detailed calculations for Type IIL events (mainly because the parameters of their progenitors are still uncertain); thus, it is a possibility that dust grains can form at even higher velocities than expected for SNe~IIP.

\vspace{3mm}

In summary, we conclude that the late-time mid-IR radiation of SN~1980K, observed serendipitously with {\it JWST}/MIRI in September 2022, can be described with a $T_\rmn{dust}\approx 150$~K, $M_\rmn{dust} \approx 0.002$~\msolar\ silicate dust component and an additional hotter dust/gas component. 
Results of analytical and numerical modeling of the mid-IR SED show that this dust can be entirely newly-formed and located in the ejecta. 
We also obtained an optical spectrum of the SN with Keck~I 10~m telescope in November 2022, showing strong emission lines of H$\alpha$ and [O~{\sc i}] with asymmetric blueshifted profiles, very similar to those seen in other spectra of SN~1980K obtained in the last several years.
Results of radiative-transfer modeling of these optical line profiles agree well with that of similar examinations published before \citep{bevan17,niculescu-duvaz22}, showing a total dust mass of $\sim 0.24$--0.58~\msolar\ presumably located in the ejecta. 
Putting these two conclusions together, we could see only part of any ejecta dust with {\it JWST}, and there should be much colder dust present.

The presence of asymmetries in the recently observed optical line profiles implies that there must be a dust component internal to the shock; however, in the mid-IR, we may see pre-existing dust heated collisionally by the SN shock wave. 
This possibility is strengthened by (i) the close similarity of the mid-IR SED of SN~1980K to the $\sim 19$~yr scaled {\it Spitzer/IRS} spectrum of SN~1987A (thought to originate from heated dust located in the ER around 87A), and (ii) the presence of the hot ($\gtrsim 400$~K) dust/gas component presumably does not connect to decades-old SN ejecta.
Well, in this case --- since mass of the primary mid-IR dust component ($T_\rmn{dust}\approx 150$~K, $M_\rmn{dust} \approx 0.002$~\msolar) is 2--3 orders of magnitudes higher than what is seen in the ER of SN~1987A, we shall imagine a kind of a ''super-ring'' (or, a different geometry/distribution of dust) in the case of SN~1980K.

Moreover, radiative heating of dust by energetic photons originating from the ongoing CSM interaction can also play a role in heating the SN dust. 
While, however, the late-time optical luminosity measured in SN~1980K is too low for heating the dust to the observed temperature alone, theoretical predictions suggest that contribution from more energetic photons could solve this discrepancy.

For a more detailed investigation, further data would be necessary: either a mid-IR spectrum to view the spectral features in detail (and, thus, to better determine the chemical composition and geometry of dust in SN~1980K), or a near-IR ($\sim 1$--5~$\mu$m) dataset for revealing the true nature of thermal radiation of the hot (dust) component, as well as UV/X-ray observations on getting deeper insight into the ongoing CSM interaction.

\section*{Acknowledgements}

We thank Richard G. Arendt for his valuable comments and for providing the {\it Spitzer}/IRS spectrum of SN~1987A.
This work is based on observations made with the NASA/ESA/CSA {\it James Webb Space Telescope}. The data were obtained from the Mikulski Archive for Space Telescopes at the Space Telescope Science Institute, which is operated by the Association of Universities for Research in Astronomy, Inc., under NASA contract NAS 5-03127 for {\it JWST}. These observations are associated with program GO-2666.
Some of the data presented herein were obtained at the W. M. Keck Observatory, which is operated as a scientific partnership among the California Institute of Technology, the University of California, and NASA; the observatory was made possible by the generous financial support of the W. M. Keck Foundation.

This project has been supported by NKFIH/OTKA grant FK-134432. S.Z. is supported by the National Talent Programme under NTP-NFTÖ-22-B-0166 Grant. T.S. is supported by the János Bolyai Research Scholarship of the Hungarian Academy of Sciences and by the New National Excellence Program (UNKP-22-5) of the Ministry for Innovation and Technology from the source of the National Research, Development and Innovation Fund. I.D.L. has received funding from the European Research Council (ERC) under the European Union's Horizon 2020 research and innovation programme DustOrigin (ERC-2019-StG-851622) and the Belgian Science Policy Office (BELSPO) through the PRODEX project “JWST/MIRI Science exploitation” (C4000142239).
A.V.F.'s supernova group at U.C. Berkeley is supported by the
Christopher R. Redlich Fund, Alan Eustace (W.Z. is a Eustace
Specialist in Astronomy), Frank and Kathleen Wood (T.G.B. is a
Wood Specialist in Astronomy), and numerous other donors.

\section*{Data Availability}

JWST photometric data underlying this article are available in the article. Spectroscopic data will be publicly released via WIseREP (\href{https://www.wiserep.org/}{https://www.wiserep.org/}).


\bibliographystyle{mnras}
\bibliography{main.bib} 

%
%
%
%

\bsp	
\label{lastpage}
\end{document}